\documentclass[a4paper,12pt]{spieman}  % use this instead for A4 paper
\usepackage{amsmath,amsfonts,amssymb}
\usepackage{graphicx}
\usepackage{setspace}
\usepackage{tocloft}

\usepackage{siunitx}
\usepackage {spverbatim}
\usepackage{listings}
\usepackage{multirow}

\makeatletter
\newcommand{\figcaption}[1]{\def\@captype{figure}\caption{#1}}
\newcommand{\tblcaption}[1]{\def\@captype{table}\caption{#1}}
\makeatother

\title{High count rate effects in event processing for XRISM/Resolve x-ray microcalorimeter{: I. Ground test}}
\author[a,*]{Misaki Mizumoto}
\author[b]{Tsubasa Tamba}
\author[b]{Masahiro Tsujimoto}
\author[c]{Renata~S.~Cumbee}
\author[d]{Megan~E.~Eckart}
\author[c]{Edmund Hodges-Kluck}
\author[e,b]{Yoshitaka~Ishisaki}
\author[c]{Caroline~A.~Kilbourne}
\author[c]{Maurice~A.~Leutenegger}
\author[c]{Frederick~S.~Porter}
\author[f]{Makoto~Sawada}
\author[g]{Yoh~Takei}
\author[h]{Yuusuke~Uchida}
\author[f]{Shin'ya~Yamada}
\affil[a]{University of Teacher Education Fukuoka, Munakata, Fukuoka 811-4192, Japan}
\affil[b]{ISAS/JAXA, Sagamihara, Kanagawa 252-5210, Japan}
\affil[c]{NASA GSFC, Greenbelt, MD 20771, USA}
\affil[d]{Lawrence Livermore National Laboratory, Livermore, CA 94550, USA}
\affil[e]{Tokyo Metropolitan University, Hachioji, Tokyo 192-0397, Japan}
\affil[f]{Rikkyo University, Toshima-ku, Tokyo 171-8501, Japan}
\affil[g]{ISAS/JAXA, Tsukuba, Ibaraki 305-8505, Japan}
\affil[h]{Tokyo University of Science, Noda, Chiba 278-8510, Japan}

\cftpagenumbersoff{figure}
\cftpagenumbersoff{table} 
\begin{document} 
\maketitle

\begin{abstract}
The spectroscopic performance of an X-ray microcalorimeter is compromised at high count rates. 
In this study, we utilize the \textit{Resolve} X-ray microcalorimeter onboard the XRISM satellite to examine the effects observed during high count rate measurements and propose modeling approaches to mitigate them.
We specifically address the following instrumental effects that impact performance: CPU limit, pile-up, and untriggered electrical cross talk. Experimental data at high count rates were acquired during ground testing using the flight model instrument and a calibration X-ray source. In the experiment, data processing not limited by the performance of the onboard CPU was run in parallel, which cannot be done in orbit. This makes it possible to access the data degradation caused by limited CPU performance.
We use these data to develop models that allow for a more accurate estimation of the aforementioned effects. To illustrate the application of these models in observation planning, we present a simulated observation of GX 13+1. 
Understanding and addressing these issues is crucial to enhancing the reliability and precision of X-ray spectroscopy in situations characterized by elevated count rates.
\end{abstract}

% Include a list of up to six keywords after the abstract
\keywords{XRISM, \textit{Resolve}, x-ray microcalorimeter, digital signal processing,
high count rate}

% Include email contact information for corresponding author
{\noindent \footnotesize\textbf{*}MM,  \linkable{mizumoto-m@fukuoka-edu.ac.jp} }

\begin{spacing}{1}   % use double spacing for rest of manuscript

\section{Introduction}\label{s1}  
An X-ray microcalorimeter is a device designed to convert the energy of incoming X-ray photons into heat, sensing the resulting temperature change in a thermometer cooled to sub-Kelvin temperatures\cite{McCammon1984}. This technology achieves unprecedented energy resolution over a wide energy range nondispersively. Placing an array of microcalorimeters at the focus of an X-ray
telescope enables collection of a high-resolution spectrum for each imaging element. Each pixel is individually anchored to the heat sink, and the signal time constant is related to the thermal time constant of that link.

The XRISM satellite\cite{Tashiro2020a} was launched on September 7th, 2023 in JST (6th in UT). At high count rates, the exceptional spectroscopic performance of the XRISM/\textit{Resolve} instrument can be affected by various factors, some of which are discussed in this paper. First, X-ray pulses can overlap due to the slow thermal time constant and even slower ac-coupling time constant. Second, for space applications, signal processing occurs onboard, and only characteristic pulse values are downlinked to fit into the telemetry bandpass\cite{Boyce1999}. Computational resources may limit processing rates, and these effects are further complicated by cross talk among multiple pixels.

The only prior example of high count rate astrophysical observation with a microcalorimeter is Crab\cite{tsujimoto18c} with the soft X-ray spectrometer (SXS)\cite{Mitsuda2014} onboard ASTRO-H\cite{Takahashi2018} in 2016. This paper aims to present a case study of these limitations and mitigation using the X-ray microcalorimeter \textit{Resolve}\cite{ishisaki2022, sato2023,kelley2024} onboard XRISM by combining experimental and simulation methods. Previous papers discussed mitigation in the optical chain, such as an X-ray attenuation filter or de-focusing mirrors\cite{Kammoun2022}, while our focus is on the signal chain.

\begin{figure}[htbp]
 \begin{center}
\includegraphics[width=160mm]{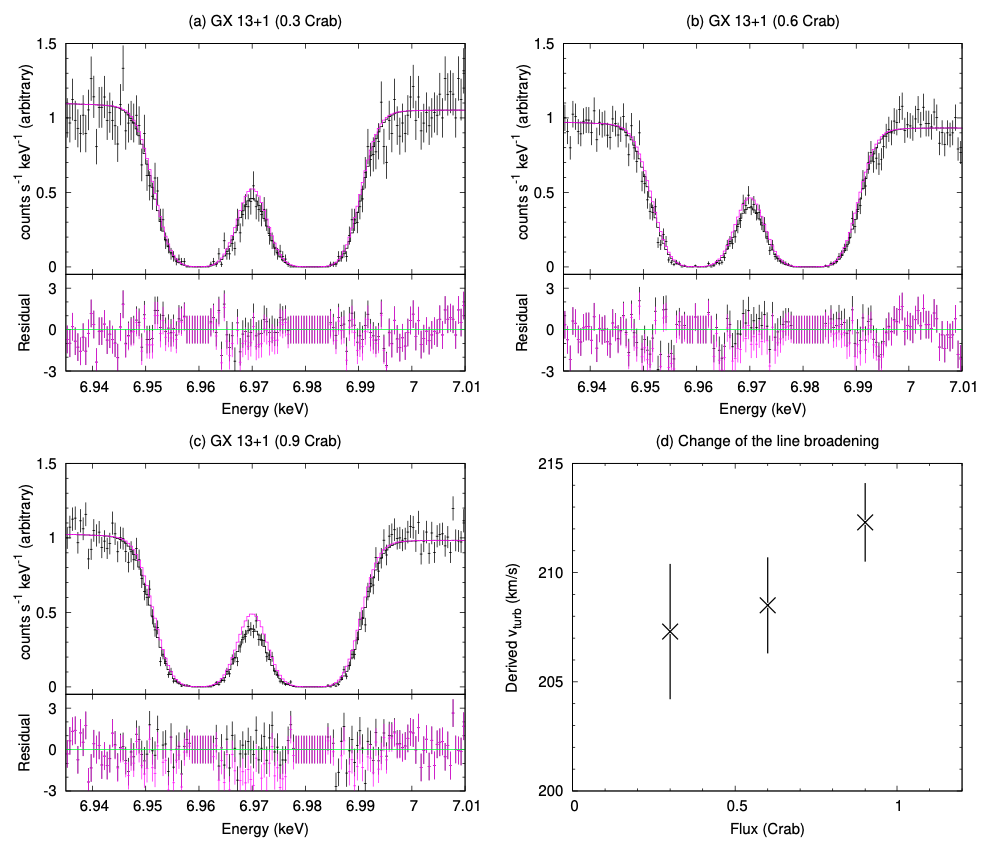}
 \end{center}
 \caption
 {\label{f01} 
(a)-(c) Simulated spectra of GX 13$+$1 {with the exposure time of 15~ks}, at three different assumed count rates. 
{The high-primary (Hp) events are simulated, with a FWHM of 5~eV.}
The vertical axis is normalized at 6.935~keV. {
If no high count rate effects exist, the observed spectrum should follow the magenta line, while the black bins depict the \textit{Resolve}-simulated spectrum with high count rate effect.} If the spectra are fitted without considering this effect (black lines), the estimation of the turbulent velocity exhibits an increasing value with the count rate, resulting in a spurious correlation between the two in (d).}
\end{figure}

We begin with Figure~\ref{f01}, where we simulate some effects of high-count-rate observations with \textit{Resolve}.
The panels show simulated X-ray spectra for GX 13$+$1, a low-mass X-ray binary (LMXB), observed by \textit{Resolve} with different count rates (or different flux). The energy resolution depends on the grade of the event, so we use events with the best energy resolution (High primary, Hp; see \S~\ref{s2} for details). We assumed an energy resolution of 5~eV in the Full-Width Half Maximum (FWHM).
The source is assumed to be located near the array center with a 1/4 Neutral Density (ND) filter in the gate valve open case.
Here, we focus on Fe Ly$\alpha^{1,2}$ absorption features with an assumed turbulent velocity of $v_\mathrm{turb}=200$~km~s$^{-1}$ 
(see \S~\ref{s4-2} for details of the fitting model). As we shall discuss in this paper, cross talk degrades energy resolution as the flux increases. If no effect exists, the simulated spectra should follow the magenta lines in Figure~\ref{f01} (a--c). However, in the \textit{Resolve} observation under the high count rate situation, the energy resolution will deteriorate, as seen in the simulated spectra (black bins in Figure~\ref{f01} a--c). When the spectra are fitted without considering any deterioration in energy resolution, one may derive a larger $v_\mathrm{turb}$ than the true value, obtain an artificial flux dependence of $v_\mathrm{turb}$ (Figure~\ref{f01} d), and draw a wrong astrophysical interpretation, such as the turbulent motion in the outflowing gas becoming larger in the brighter phase.

For proper observation planning and data analysis, these behaviors need to be described and modeled based on actual data taken under a controlled setup, which will be discussed in this paper with the following plan. In \S~\ref{s2}, we provide a brief description of the \textit{Resolve} instrument, focusing on topics relevant to this study. In \S~\ref{s3}, we discuss the experimental part of this work, focusing on three effects in the signal chain: (1) CPU limit, (2) pulse pile-up, and (3) spectral degradation by electrical cross talk. The data acquisition (\S~\ref{s3-1}) and analysis (\S~\ref{s3-2}) using the hardware and software for flight are shown, and the experimental results of event loss and spectral distortion are discussed (\S~\ref{s3-3}).
The simulation is shown in \S~\ref{s4}, first for modeling these effects (\S~\ref{s4-1}) and presenting a case study in astrophysical application (\S~\ref{s4-2}). 

{
In this paper, we focus on the results of the ground test. 
In orbit, we can only obtain data that is limited by the processing capabilities of the onboard CPU. On the other hand, in this work, as explained in \S~\ref{s3-1} in detail, we have set up a unique configuration that connects a device with almost no  limitations  on data processing in parallel with the onboard CPU, allowing us to simultaneously obtain data without the limitation.
By comparing the limited data with the non-limited data, we gain insight into the non-linear effects induced by high count rates. In addition to this, we can control the X-ray brightness that illuminates the detector, allowing for a comprehensive performance evaluation under varied conditions. These capabilities make this work an indispensable complement to in-orbit observations, ensuring the accuracy and reliability of XRISM’s high-resolution X-ray spectroscopy.
The in-orbit results for the high count rate situation (e.g., Crab, Cyg X-1) will be shown in our subsequent paper.}

Throughout this paper, {we use a conversion of 1~mCrab = 2.14~s$^{-1}$~array$^{-1}$, in which the Crab flux and spectrum\cite{willingale2001} are assumed as a point-like source placed at one of the center pixels with an open filter.} This conversion is used for contextualizing the hardware limitations in astrophysical terms. Here, we assume the gate valve open case; the study for the gate valve closed case will be presented in the subsequent paper. 
%Thermal cross talk\cite{kilbourne18b}, which is much smaller than electrical cross talk, is not considered in this paper.
Error bars show $1\sigma$ statistical uncertainty unless otherwise mentioned.

%%%%%%%%%%%%%%%%%%%%%%%%%%%%%%%%%%%%%%%%%%%%%%%%%%%%%%%%%%%%%%%%%%%%%%
\section{Instrument}\label{s2}
%%%%%%%%%%%%%%%%%%%%%%%%%%%%%%%%%%%%%%%%%%%%%%%%%%%%%%%%%%%%%%%%%%%%%%
The X-ray microcalorimeter for \textit{Resolve} is based on ion-implanted Si thermometers with HgTe X-ray absorbers anchored to a 50~mK heat sink with a thermal time constant of $\sim$3.5~ms. The array consists of 6$\times$6 pixels (PIXEL 0--35)\cite{kilbourne18b, porter18}, below which an anti-coincidence detector\cite{kilbourne18b, porter18} (anti-co) with two readouts is placed for identifying cosmic-ray events. 
One of the 36 pixels (PIXEL 12) is displaced from the array to be illuminated by an $^{55}$Fe source for tracking gain changes common to the entire array. The thermometer resistance of $\sim$30~M$\Omega$ is read individually for each pixel, which is converted to low impedance with junction field effect transistor (JFET) amplifiers to go outside of the cryostat. The JFET operates at 130~K, thus needs to be physically separated from the 50~mK stage\cite{Kelley2016}.
The electrical cross talk takes place between adjacent channels in the high-impedance part of the circuit prior to the JFET trans-impedance amplifiers. The pixels are enumerated by the wire layout of this part, so pixel $i$ cross-talks to pixels $i\pm1$ with a peak pulse height of 0.6\% (relative physical pulse height before optimal filtering), and $i\pm2$ with 0.1\% in the same detector quadrant. {Sequential groups of nine pixels define the quadrants (PIXEL 0--8, 9--17, 18--26, and 27--35).}

\begin{figure}[htbp]
 \begin{center}
\includegraphics[width=160mm]{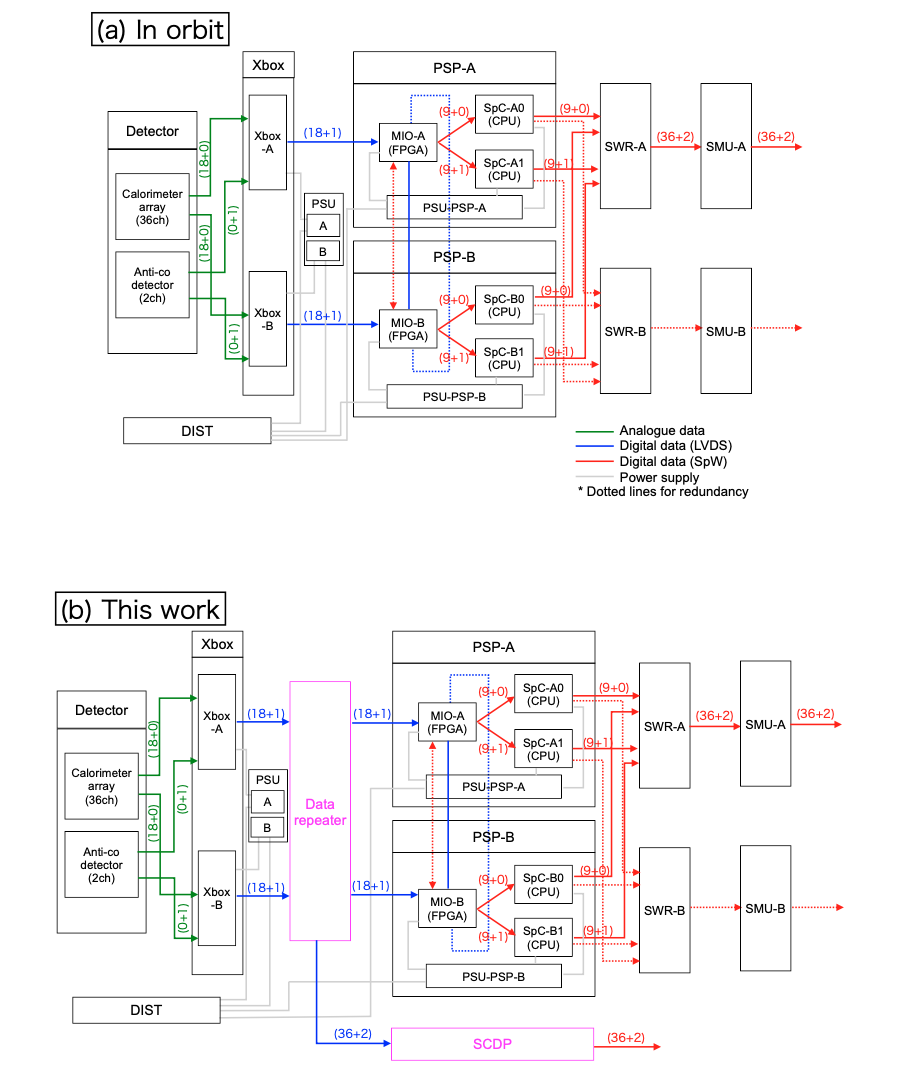}
  \caption
  { \label{fig:dataflow} 
  Component diagram and routes for event data in orbit (upper) and in this work (lower). The numbers in parentheses indicate the number of pixels (calorimeter array + anti-co) for which the route is responsible. The difference between (a) and (b) is highlighted in magenta. In this work, the data repeater is inserted between Xbox and PSP, and the event data is also processed in Software Calorimeter Digital Processor (SCDP; see \S\ref{s3-1}) in parallel with PSP. MIO=Mission Input/Output board, SpC=Space Card, SpW=Space Wire, and LVDS=Low Voltage Differential Signaling.
  }   
 \end{center}
\end{figure}

Signals are ac-coupled, inverted, amplified, and sampled at 12.5~kHz with a bipolar 14~bit analog-to-digital converter (ADC) in the onboard analog electronics (Xbox)\cite{Kelley2016}, {which receives regulated and stable power from a standalone Power Supply Unit (PSU).
The digitized signal} is relayed to the onboard digital electronics (Pulse Shape Processor; PSP)\cite{ishisaki18a} for event detection and reconstruction. The PSP consists of two identical units (PSP-A and PSP-B), each of which has one Field Programmable Gate Array (FPGA) board, two CPU boards, and one {power supply board (PSP-PSU).}
The FPGA boards are responsible for event candidate detection, while the CPU boards are for identifying overlapping events, grading them, and applying optimal filtering. Each of the four CPU boards (PSP-A0, A1, B0, and B1) processes one quadrant{: PSP-A0 is responsible for processing the data from PIXEL 0--8, A1 for PIXEL 9--17 and one anti-co channel (anti-co A), B0 for PIXEL 18--26, and B1 for PIXEL 27--35 and the other anti-co (anti-co B)}. All channels are processed in parallel, and all cross-channel processing is applied on the ground.
The digital data processed by PSP is relayed to the data recorder via the SpaceWire Router (SWR) and the Satellite Management Unit (SMU). The SWR and SMU are composed of two identical units for redundancy. The power for {the PSP and the separate PSU supplying the Xbox} is provided via the power distributor (DIST).
Figure \ref{fig:dataflow} (a) shows the component diagram of these devices in orbit.

The CPU board utilizes an SH4-compatible 32-bit Reduced Instruction Set Computer (RISC) processor, operating at a clock speed of 60 MHz, developed by Mitsubishi Heavy Industries. The real-time operating system used is \texttt{TOPPERS}.
When the incoming event rate is higher than $\sim50$~s$^{-1}$~quadrant$^{-1}$, the CPU in the PSP has difficulties processing all events in that quadrant, leading to event loss. In such cases, corrections are needed to accurately calculate the flux, and it may not be possible to identify all events contaminated by the cross talk (refer to \S\ref{s3-3-3}). Here, we briefly outline the mechanism of event loss (for a more detailed explanation, see Appendix \ref{sec:PSPbuffer}).

Event candidates are triggered by the FPGA and then processed by the CPU in PSP, as illustrated in Figure \ref{fig:dataflow}. These events are temporarily stored in buffers while awaiting CPU processing. If the rate at which event candidates arrive (i.e., the incoming rate to the buffer) exceeds the CPU's maximum event processing capacity (i.e., the maximum output rate from the buffer), the number of events in the buffer will increase. If this is only a temporary surge, the CPU may be able to catch up before the buffer becomes full. However, if event candidates continue to arrive at a high rate, the buffer may reach its limit and discard its contents at once, which results in event loss. In this paper, it is called PSP overflow.

Each CPU processes event candidates from a specific quadrant, and each pixel has its own buffer. This means that when the CPU in a quadrant operates at full capacity (100\% consumption rate), some pixels with higher count rates may experience significant event loss, while other pixels with lower count rates might not experience any event loss. This is because the real-time operating system processes events using a round-robin scheduling method, which allocates as equal processing time as possible to each pixel regardless of its count rate. Therefore, it is crucial to check for event loss pixel-by-pixel.

The CPU consumption rate is recorded in the telemetry file, making it accessible to users. If event loss occurs, PSP records the pixel where the loss happened and the start and end times of the event loss (referred to as the bad time interval). By applying the bad time interval to each pixel, it is possible to correct the ``true'' exposure time and accurately restore the incoming event rate.

The CPU consumption rate is a non-linear function of the count rate. Events are graded into High, Mid, and Low resolution (HR, MR, and LR) depending on how close they are in time to an adjacent event of the same pixel\cite{ishisaki18a}. They are further subdivided into primary (p) or secondary (s) for those without and with a preceding event. For HR ($=$Hp) / MR ($=$Mp$+$Ms) events, the template of a 1024/256 sample length (81.92/20.48~ms) is used for optimal filtering to derive the characteristic values of an event, including arrival time, energy, \texttt{DERIV\_MAX}, and \texttt{RISE\_TIME}. \texttt{DERIV\_MAX} is the maximum value of the derivative, and \texttt{RISE\_TIME} is the time in units of 20~\si{\micro s} from \texttt{DERIV\_MAX} to the zero-crossing of the derivative, corresponding to the rise time of the raw pulse. For LR ($=$Lp$+$Ls) events, only coarse values of these characteristics are derived based on the FPGA. As the count rate increases, the resource-demanding Hp events decrease and the less-demanding Ls events increase.

The gain of an x-ray calorimeter is sensitive to its environment including heat sink temperature, bolometric loading, and the temperature of the readout electronics. To correct the time dependent gain on orbit, we utilize a non-linear temporal gain correction technique developed for Astro-H, XRISM, and Athena/XIFU\cite{porter16, porter2024,cucchetti2018,smith2023,cucchetti2024}.  For XRISM, the technique essentially parametrizes all time dependent gain fluctuations as an ``effective'' pixel temperature since this is the dominant mechanism affecting the gain. We then use an on-board fiducial source to estimate the effective pixel temperature for each x-ray event. With the XRISM/Resolve gate-valve closed, we use a set of $^{55}$Fe fiducial x-ray sources mounted on the instrument filter wheel (FW) to provide the gain fiducial. The FW sources are periodically rotated into the Resolve aperture to monitor the detector gain. We then synthesize a new non-linear energy-scale curve for each event in the observation by interpolating a set of temperature and pixel dependent energy scale curves measured during ground calibration and verified in flight\cite{eckart2024}. This process has been demonstrated to be extremely effective in recovering the energy scale in flight to well within the $\pm2$~eV requirement, typically better than 0.2~eV across the XRISM/Resolve passband\cite{porter2024, eckart2024}.

%%%%%%%%%%%%%%%%%%%%%%%%%%%%%%%%%%%%%%%%%%%%%%%%%%%%%%%%%%%%%%%%%%%%%%
\section{Experiments}\label{s3}
%%%%%%%%%%%%%%%%%%%%%%%%%%%%%%%%%%%%%%%%%%%%%%%%%%%%%%%%%%%%%%%%%%%%%%
\subsection{Data Acquisition}\label{s3-1}
\begin{figure}[htbp]
 \begin{center}
  \includegraphics[width=120mm]{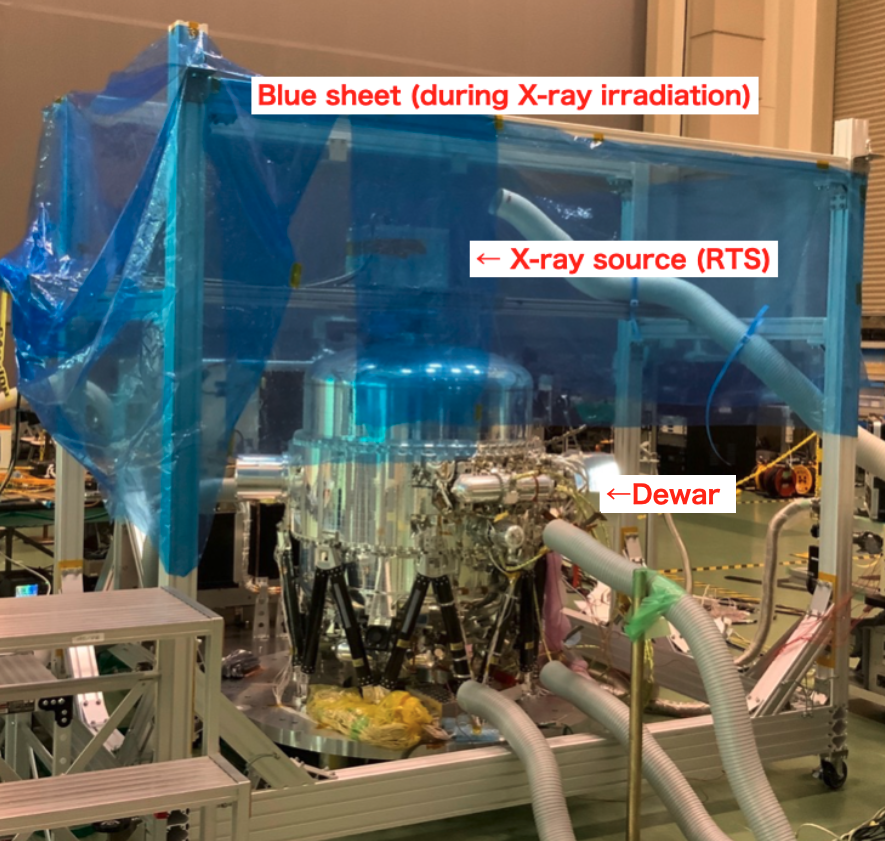}
  \caption
  { \label{fig:dewar} 
  Photo of the experimental setup.
  }   
 \end{center}
\end{figure}

We obtained high count rate data during the instrument-level test at the Tsukuba Space Center of JAXA on February 20--21, 2022 (Figure \ref{fig:dewar}). The flight model hardware and final optimal filtering templates were utilized. The instrument was operated in cryogen-free mode. 
We anticipate that the results will exhibit similar characteristics in both cryogen mode and cryogen-free mode.
The instrumental setting for the signal chain in this work is almost the same as the flight setting, except for the insertion of the data repeater between Xbox and PSP (Figure \ref{fig:dataflow} b). 
All incoming data from the Xbox are streamed to a personal computer (PC), in parallel with the nominal data flow to the PSP. On the PC, PSP-equivalent processing, but with virtually no CPU limits, was run using the Software Calorimeter Digital Processor (SCDP). 
Therefore, we were able to process the data in the same way as in flight and also, in parallel, without CPU limitation.
This test was performed at the highest level of integration with the data repeater. Similar data cannot be taken after this experiment, either on the ground or in orbit.

\begin{figure}[htbp]
 \def\@captype{table}
 \begin{minipage}[t]{.48\textwidth}
  \begin{center}
   \begin{tabular}{cccc} 
    \hline
    Step & Target & Current & Exposure  \\
    & & (\si{\micro \ampere}) & (min)  \\
    \hline
    1&Ni & 12 & 10 \\
    2&& 200	& 30 \\
    3&& 12	&10 \\
    4&& 131	&  30 \\
    5&& 12	&  10 \\
    6&& 102	& 10 \\
    7&& 88 & 10 \\
    8&& 12  & 10 \\
    9&& 73 & 10 \\
    10&& 58  & 10 \\
    11&& 12  & 10 \\
    12&& 44 & 10 \\
    13&& 29  & 10 \\
    14&& 12  & 10 \\
    15&& 9  & 20 \\
    16&& 5 & 20 \\
    17&& 12  & 10 \\
    \hline
    18&KBr & 12 &10 \\
    19&&  120 &30\\
    20&&  12  &10\\
    21&&  6 &15 \\
    \hline
   \end{tabular}
  \end{center}
  \tblcaption{Experimental setup in the uniform illumination test for {each experimental sequence (``Step'' in the table)}.}
  \label{tab:obs}
 \end{minipage}
 \hfill
  \begin{minipage}[t]{.48\textwidth}
  \begin{center}
   \begin{tabular}{cccc} 
    \hline
    Step & Anode target & Current & Exposure  \\
    & & (\si{\micro \ampere}) & (min)  \\
    \hline
    22&Cu & 10 & 10 \\
    23&& 100	& 30 \\
    24&& 10	& 10 \\
    25&& 50	& 20 \\
    26&& 10	& 10 \\
    27&& 30	& 10 \\
    28&& 10	& 10 \\
    29&& 150	& 30 \\
    30&& 10	& 10 \\
    \hline
    \multicolumn{4}{c}{(Move spot to a corner of detector)}\\
    31&Cu & 10 &10 \\
    32&&  100 &30\\
    33&&  10  &10\\
    34&&  50 &20 \\
    35&&  10  &10\\
    36&&  30 &20 \\
    37&&  10  &10\\
    \hline
   \end{tabular}
  \end{center}
  \tblcaption{Experimental setup in the point-like illumination test for {each experimental sequence.}}
  \label{tab:obs2}
 \end{minipage}
\end{figure}

We conducted our experiment in two different settings: a uniform illumination test and a point-like illumination test. {The former has richer statistics, while the latter simulates observations of bright compact sources, such as X-ray binaries.}

In the uniform illumination test, we utilized X-ray sources developed for the ground calibration of the gain curve (RTS; rotating target source)\cite{eckart18}. X-rays were generated by the fluorescence of a selected target, illuminating uniformly across the array.
%Note that this setup gave a much higher total count rate than that in actual in-orbit observations, causing some thermal loading.  The results from this experiment are, however, valid except for the gain shift described in Appendix \ref{sec:energyshift}.
{The high voltage current was
adjusted to obtain different rates} in the $\sim$0.4--15.3~s$^{-1}$~pixel$^{-1}$ range for a 10--30~min exposure (Table \ref{tab:obs}). This incoming rate covers CPU usage from low rates to 100\%.
In this experiment, measurements were performed under conditions with an exceptionally high count rate. To estimate the impact of rapid count-rate fluctuations on gain shift, we utilized low count-rate fiducials (the steps with 12~\si{\micro \ampere} in Table \ref{tab:obs} and those with 10~\si{\micro \ampere} in Table \ref{tab:obs2}) inserted both before and after each high-count-rate step in the experiment, similar to the method used in flight. 

\begin{figure}[htbp]
 \begin{center}
  \includegraphics[width=150mm]{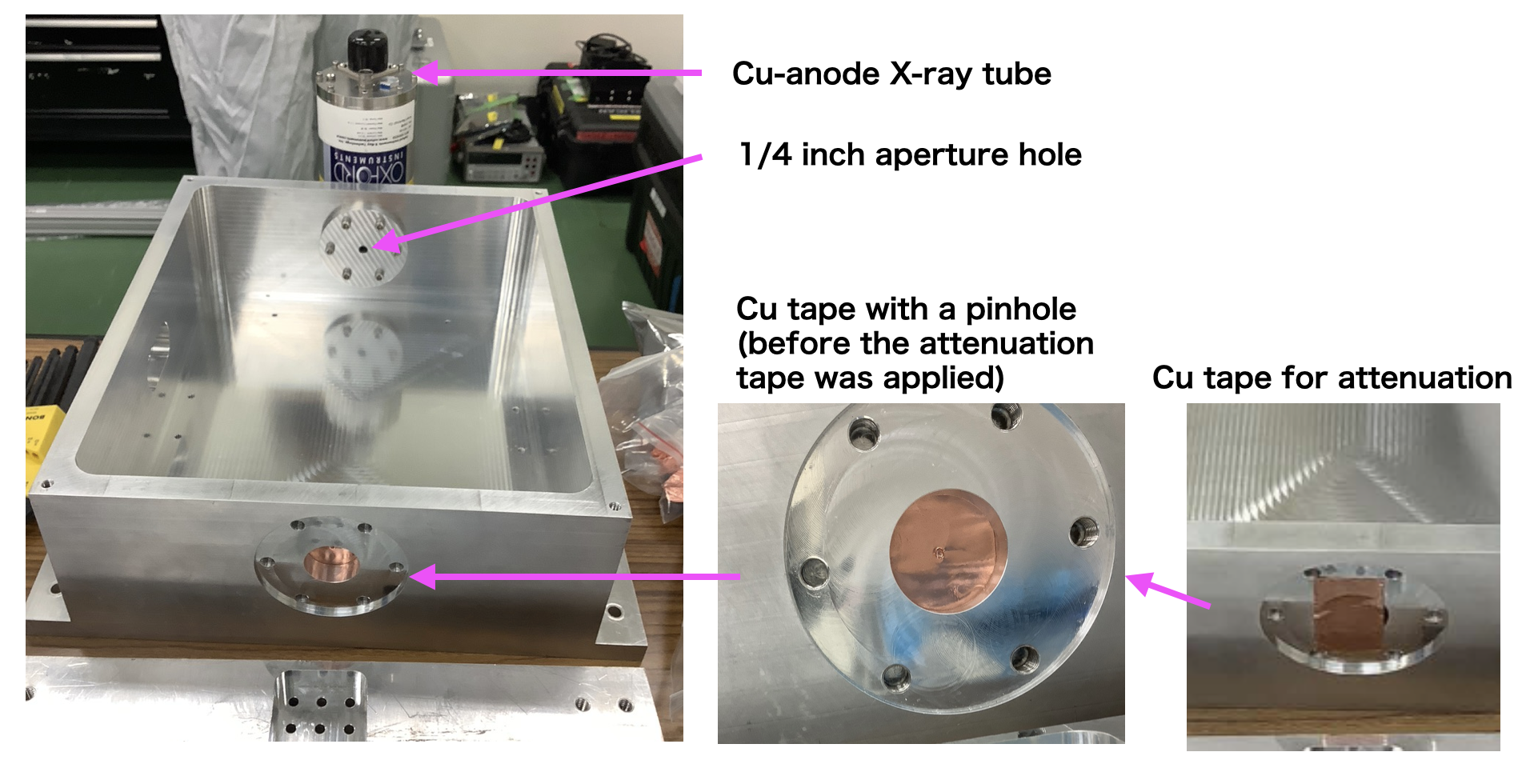}
  \caption
  { \label{fig:cutape} 
The setup for the point-like illumination test. Cu florescent X-rays produced by the Cu-anode X-ray tube are narrowed by passing through a pinhole on the copper tape and spotlighted on the array, with the bremsstrahlung continuum. The number of counts in the Cu lines is $\sim27$~\% of the total one.
  }   
 \end{center}
\end{figure}

\begin{figure}[htbp]
 \begin{center}
\includegraphics[width=160mm]{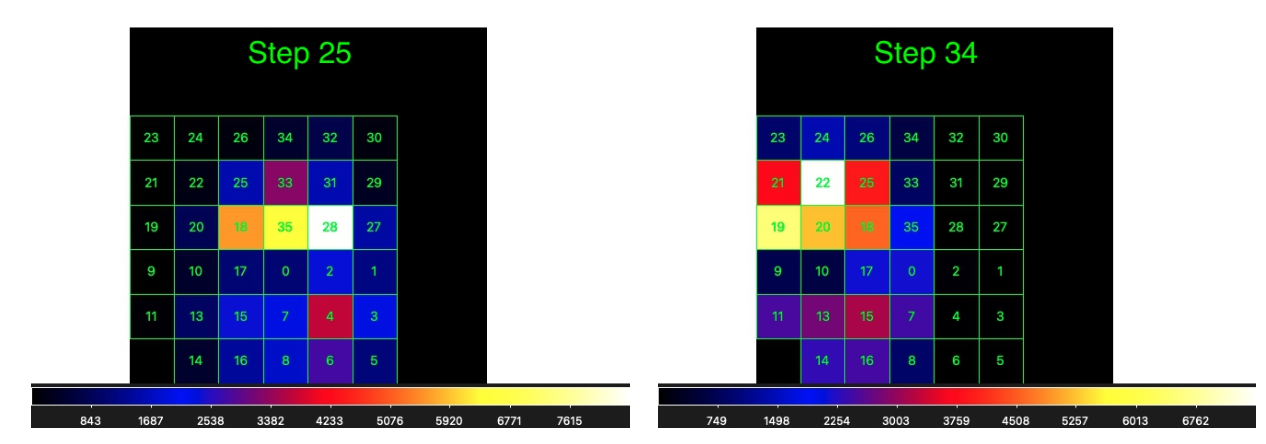}
 \end{center}
 \caption
 {\label{fig:ds9} 
 The X-ray count map in the point-like illumination test. The number of Hp events per pixel during each step is shown. The green numbers show the pixel numbers. The left panel shows the one in Step 25 (when the bright spot is located at the array center) and the right panel shows Step 34 (when the bright spot is near the corner of the array). In both cases the current was low and thus no PSP overflow occurred.}
\end{figure}

In the point-like illumination test, we used Cu fluorescent X-rays emitted from the Cu-anode X-ray tube (Oxford Series 5000). A collimating aperture was added at the exit of the aluminum housing to shape the beam, simulating a point source. The aperture consists of two parts: the first part has 4 layers of 75~\si{\um} Cu tape with a hole of 1.5~mm in diameter, and the second part has 1 layer of 75~\si{\um} Cu tape with a hole of 0.5~mm in diameter. Behind the aperture, two layers of 75~\si{\um} tape without holes were introduced for attenuation of flux. The setup is shown in Figure \ref{fig:cutape}. 
{Maximum processed rate for the quadrant will depend on the distribution of rates among the pixels in the quadrant. To see this effect, we have two sets of experiments with different count maps.
That is, }the bright spot was initially located close to the center of the array and then moved to near the corner (Figure \ref{fig:ds9}). {The experimental setting is listed in Table \ref{tab:obs2}.} The incoming count rate is in the range of 0.4--16.9~s$^{-1}$~pixel$^{-1}$  at the pixel with the highest rate. We only used pixels with sufficient photon counts, specifically PIXEL 18, 28 and 35 in Steps 22--30 and PIXEL 18--22 in Steps 31--37. The low count rate fiducials were also inserted.

\subsection{Data Reduction \& Analysis}\label{s3-2}
\subsubsection{Data screening}\label{s3-2-1}
Data were processed using the XRISM pre-pipeline\cite{Terada2021} and pipeline\cite{Loewenstein2020} tools, in the same way as for flight data.
We made some modifications to ensure compatibility with the instrument-level test data. 
The standard screening criteria used in ASTRO-H SXS\cite{hitomi_anaguide} were applied to make the ``cleaned'' events, unless otherwise mentioned. The uniform illumination test data, with richer statistics than those of the point-like illumination test, were used for all assessments (\S\ref{s3-3-1}--\S~\ref{s3-3-3}), while the point-like illumination test data were used solely for spectral analysis (\S~\ref{s3-3-3}).

\subsubsection{Gain correction}\label{s3-2-2}
\begin{figure} [htbp]
 \begin{center}
  \begin{tabular}{cc} 
 \begin{minipage}[t]{80mm}
    \centering
    \scalebox{0.5}{\includegraphics{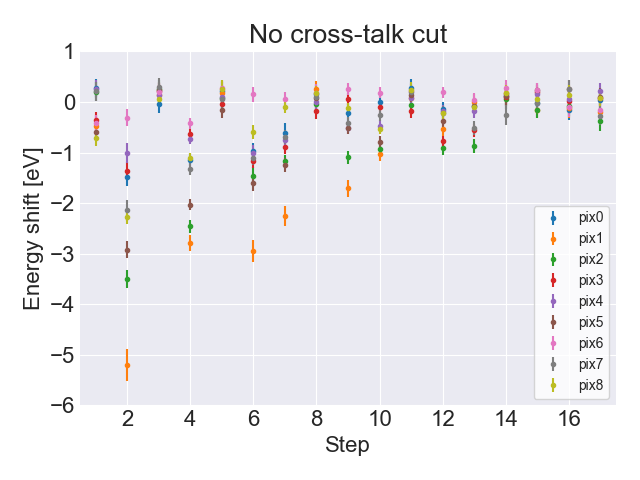}}
 \caption
 {\label{fig:gainshift2} 
   Energy shift at the Ni K$\alpha$ line in the uniform illumination test, after the standard time dependent gain correction has been applied. Only the results for Pixels 0 to 8 are shown. The cross-talk cut discussed in \S\ref{s3-3-3} has not been applied.}
  \end{minipage} 
   &
   \begin{minipage}[t]{80mm}
 \centering
    \scalebox{0.5}{\includegraphics{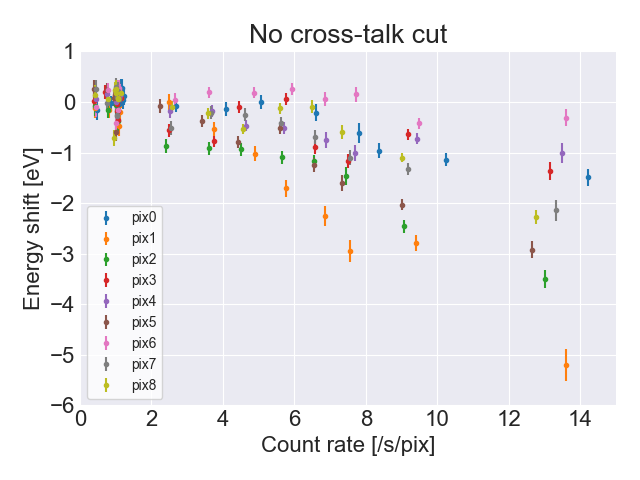}}
\caption
 {\label{fig:gainshift3} 
  Same as Figure \ref{fig:gainshift2}, but the horizontal axis shows the incoming rate. The energy shift becomes more negative as the incoming rate increases.}
  \end{minipage} \\ 
   \begin{minipage}[t]{80mm}
       \centering
    \scalebox{0.5}{\includegraphics{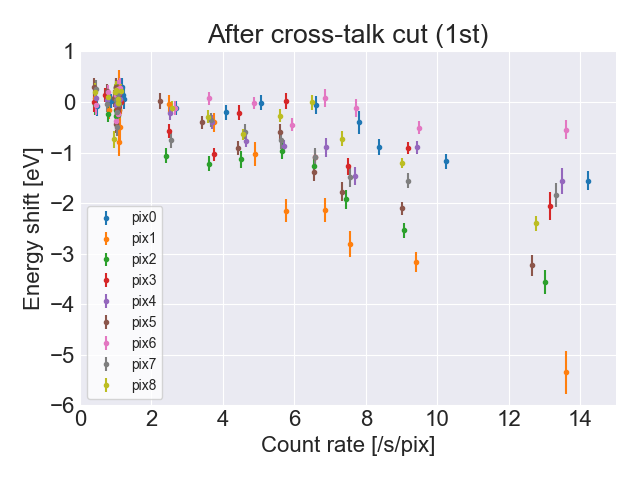}}
 \caption
 {    \label{fig:gainshift4} 
   Same as Figure \ref{fig:gainshift3}, but with the cross-talk cut of the electrical first neighbor. The negative trend in Figure \ref{fig:gainshift3} is still seen.}
  \end{minipage} 
   &
   \begin{minipage}[t]{80mm}

  \end{minipage} \\ 
  \end{tabular}
 \end{center}

\end{figure}

As explained in \S\ref{s3-1}, low-count rate fiducials were inserted both before and after each high-count rate step.
Figure \ref{fig:gainshift2} provides an evaluation of the energy shift using the low-count-rate fiducials. The line energy shifts to lower energy in the high-count-rate steps. Figure \ref{fig:gainshift3} shows the gain shift as a function of the incoming rate in each pixel; the gain shifts to lower energy as the incoming rate increases. This trend remains essentially unchanged after a cross-talk cut (Figure \ref{fig:gainshift4}), as discussed in \S\ref{s3-3-3}. This shift is likely due to local heating of the detector frame between and around the pixels in the detector array. This affects the heat sink temperature for each pixel in the array. In the uniform illumination test, some X-rays illuminated the detector frame outside the main detector array and may cause additional heating effects. As an example, the Ni line with 7.47~keV ($=1.2 \times 10^{-15}$~J) and with the count rate of 14 cts s$^{-1}$ pix$^{-1}$ illuminates the array. Assuming the illuminated area is three times the area of 36 pixels, which is reasonable considering the aperture geometry, the full power is $1.2\times10^{-15}\,\mathrm{J}\times14\,\mathrm{cts\,s}^{-1}\,\mathrm{pix}^{-1}\times36\,\mathrm{pix}\times3=1.8\times10^{-12}$~W. With the thermal conductance of the whole array
to the heat sink of $G=(6-8)\times10^{-7}$~W~K$^{-1}$ \cite{kilbourne18b},
the temperature shift in the heat sink would be $\sim2-3$~\si{\micro K}. 
In the Mn K$\alpha$ enery band (5.9 keV), the 1~mK change of the detector temperature corresponds to $\sim300$~eV\cite{porter2024}, so that the $\sim2-3$~\si{\micro K} change corresponds to $\sim0.6-0.9$~eV.
This value approximately matches the $\sim2$~eV gain shift at 7.5~keV we measured in this experiment. The verification of this assumption using in-orbit calibration observations will be reported in a separated paper. To correct the remaining energy shifts in the measurements reported here, we individually fit each high-count-rate spectral line for each pixel and count-rate interval and then adjusted the gain to recover the correct energy scale.
We simply use a linear gain stretch here since we only need the local energy scale to be correct for the measurements reported here.

\subsection{Results}\label{s3-3}
\begin{figure}[htbp]
 \def\@captype{table}
\begin{minipage}[t]{.48\textwidth}
  \resizebox{\textwidth}{!}{\includegraphics{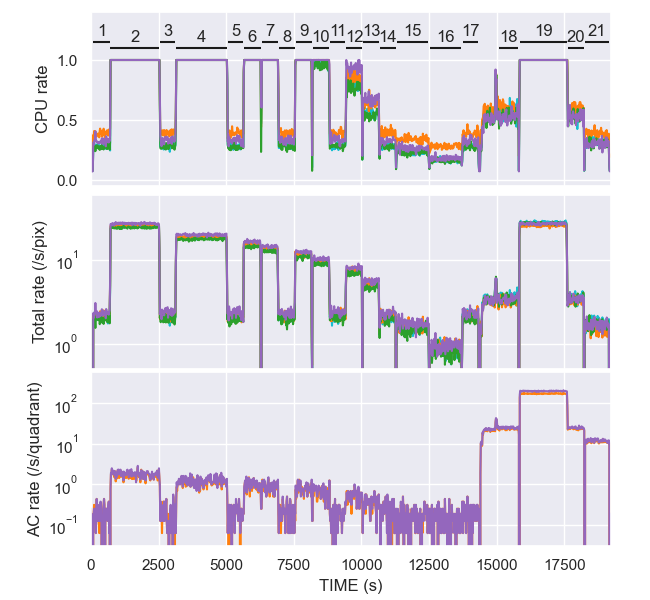}}
  \caption{
  \label{fig:PSPlimit}
  (from top to bottom) Rate of CPU consumption, the count rate of FPGA-detected event candidate with all the grades averaged in the quadrant, and the count rate of anti-co events.
  Cyan, orange, green, and purple show PSP-A0, A1, B0, and B1, respectively.  
  The step IDs in Table \ref{tab:obs} are shown at the top.
  }
 \end{minipage}
 \hfill
 \begin{minipage}[t]{.48\textwidth}
  \resizebox{\textwidth}{!}{\includegraphics{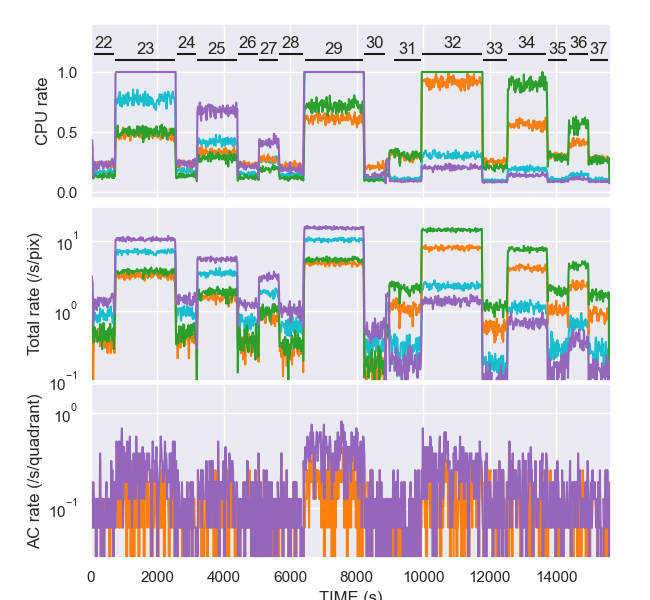}}
  \caption{
  \label{fig:PSPlimit2}
  Same as Figure \ref{fig:PSPlimit}, but for the point-like illumination test. The step IDs in Table \ref{tab:obs2} are shown at the top. 
  }
 \end{minipage}
\end{figure}

Figures \ref{fig:PSPlimit} and \ref{fig:PSPlimit2} show the CPU usage, history of the effective temperature, and the count rates.
The CPU usage changes with stepwise changes 
of count rate, and the PSP overflow occurs in some steps.
The second panel shows the count rate of FPGA-detected event candidates with all the grades, which is free from the CPU limit. This information is recorded in the telemetry files, so that users can access this information even without the SCDP, that is, even in orbit.
The last panel shows the anti-co count rate per quadrant. When the signal is uniformly illuminated, the anti-co rate becomes larger than the point-like test. In Figure \ref{fig:PSPlimit}, the KBr illumination has a higher energy (11.9 keV) than the Ni line (7.5~keV), so that more photons transmitted through the HgTe absorbers and reached the anti-co detector. This information will be used in the CPU consumption rate modeling in \S\ref{s4-1-1}.

\subsubsection{CPU Limit}\label{s3-3-1}
\begin{figure} [htbp]
 \begin{center}
  \includegraphics[width=140mm]{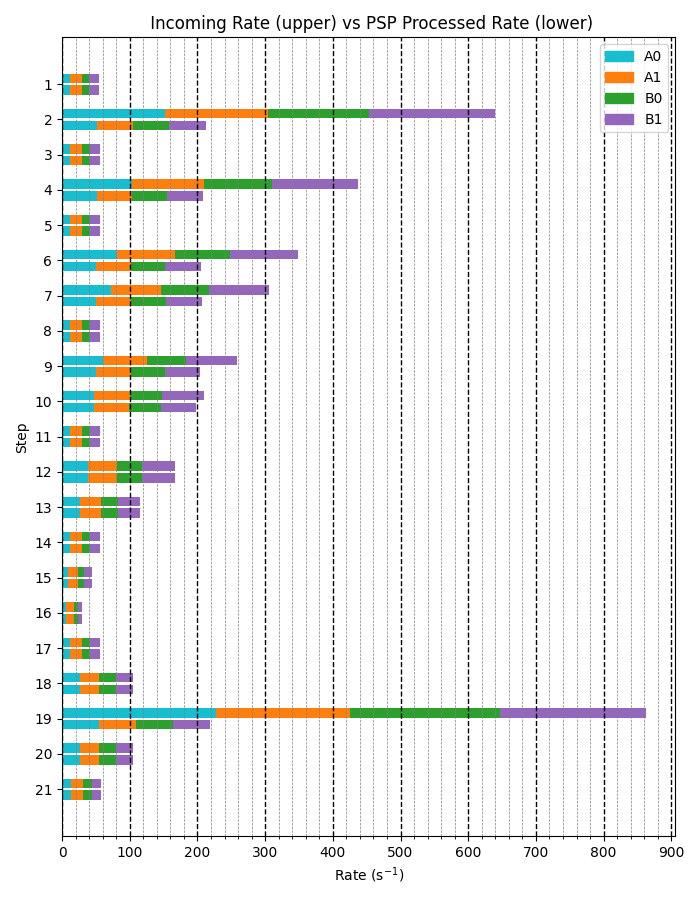}
  \caption
  { \label{fig:unfiltered} 
  Total incoming event rate (upper) and PSP-processed rate (lower) in each quadrant and each step. When the incoming rate exceeds $\sim50$~s$^{-1}$ quadrant$^{-1}$, event loss occurs and the PSP-processed rate becomes smaller than the incoming rate, while they share the same value in the other steps.
  }    
 \end{center}
\end{figure}

\textit{Resolve} has a requirement to process a count rate of $>$200~s$^{-1}$~array$^{-1}$ (or $>$50~s$^{-1}$~quadrant$^{-1}$), including background and spurious events. 
We first verify that \textit{Resolve} meets this requirement.
The objective here is to show that the PSP is able to process at least 50~s$^{-1}$~quadrant$^{-1}$ events even when the incoming rate is high enough that event loss occurs.
%The processing capability of the PSP has a limit; the processed count rate in the PSP reaches its maximum when the CPU consumption rate is 100\%.
{Figure} \ref{fig:unfiltered} shows the  total PSP-processed count rate and incoming count rate.
{The event loss and PSP overflow occur in all quadrants in Step 2, 4, 6, 7, 9, and 19, and on A1 and B1 in Step 10. 
Even in the quadrant with the PSP overflow, the PSP-processed rate  always exceeds 50~s$^{-1}$~quadrant$^{-1}$, which demonstrates that the requirement is met.
It should be noted that,} in this dataset, about 76\% of them are cleaned events suitable for astrophysical use, whereas the others are noise, triggered cross-talk events, etc.

\begin{figure} [htbp]
 \begin{center}
\includegraphics[width=160mm]{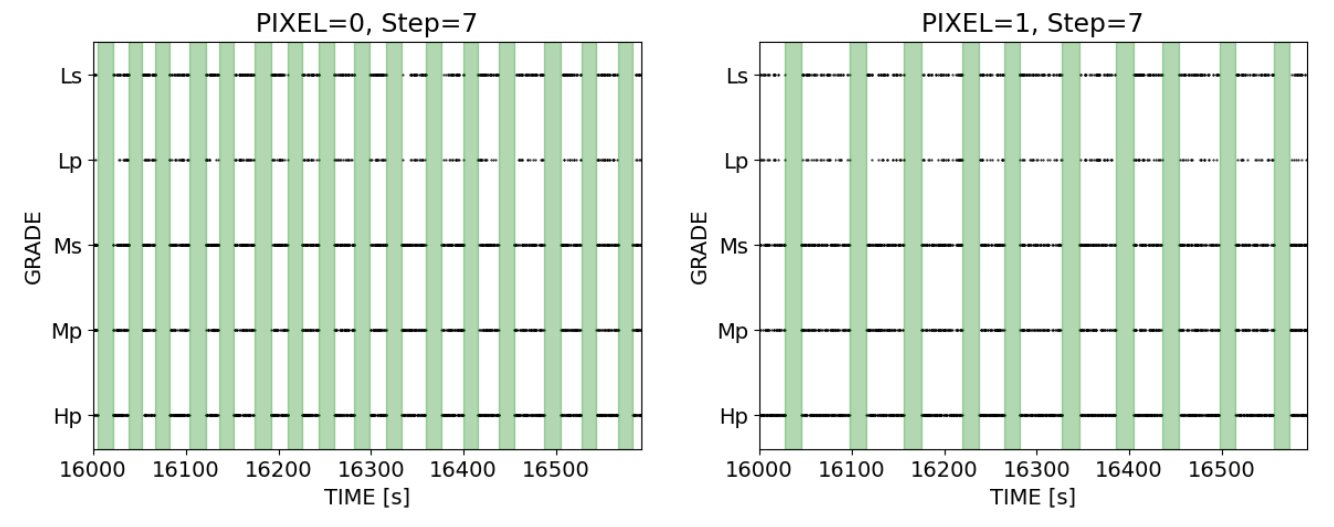}
 \end{center}
 \caption
 {\label{fig:ltcrv} 
 Example of the bad time interval in PSP. Two pixels (0 and 1 with a corrected count rate of 7.45 and
 7.94~s$^{-1}$~pixel$^{-1}$, respectively) in the step 7 in Table~\ref{tab:obs} have
 bad time intervals shown with green shades, during which all events are discarded. These intervals are listed in the event loss telemetry file. Dots show the
 grade and the arrival time of each event.}
\end{figure}

We then investigate whether the incoming rate beyond the CPU limit can be corrected using the lost event telemetry. Figure~\ref{fig:ltcrv} shows an example of the bad time intervals (intervals when the PSP did not process the events) for two selected pixels (0 and 1) in Step 7, each with an incoming rate of 7.45 and 7.94~s$^{-1}$~pixel$^{-1}$ and a bad time fraction of 45\% and 31\%, respectively. 

The actual count rate can be derived by compensating for the bad time, which can be derived with the lost event telemetry. The pixel count rates thus corrected  are summed for the array count rate, which is compared to the SCDP result in Table~\ref{tab:countrate}. 
Except for Steps 2 and 19, where the count rate is particularly high, the corrected count rate (Column 6 in Table \ref{tab:countrate}) is consistent with the SCDP (i.e., incoming rate, Column 7); the count rate is recovered. The corrected rates in Steps 2 and 19 are short of a full recovery by a maximum of 4\%. We note that this is due to the pile-up effect, as described in the next subsection (\S\ref{s3-3-2}).

\begin{table}[htbp]
 \caption{{Rate of the PSP-processed events for PIXEL 0} (1=HR, 2=MR, 3=LR, 4=Total, 5=cleaned events total,
 6=cleaned events total corrected for the bad time interval),  and SCDP processed
 rate (7). 1--(5)/(6) shows bad time percentage due to the PSP overflow, and (6)/(7) shows the efficiency due to the pile-up effect.}
 \label{tab:countrate}
 \begin{center}       
  \begin{tabular}{ccccccccccc}
   \hline
  \multirow{2}{*}{Target} & \multirow{2}{*}{Step} &  \multicolumn{6}{c}{PSP processed rate} & \multirow{2}{*}{{1$-$(5)/(6)}} & SCDP processed rate & \multirow{2}{*}{(6)/(7)} \\ 
  & & (1) & (2) & (3) & (4) & (5)  & (6)  & &(7) & \\
   \hline
  Ni & 2&0.46 & 2.64 & 2.59 & 5.69 & 4.59 & 14.75 & 0.69 &15.32 &0.96\\
   &4&0.90 & 2.77 & 1.96 & 5.63 & 4.40 & 10.43 & 0.58 &10.43 &1.0\\
   &6&1.40 & 2.52 & 1.59 & 5.51 & 4.46 & 8.49 & 0.48 &8.52 &1.0\\
   &7&1.40 & 2.51 & 1.56 & 5.48 & 4.07 & 7.45 & 0.45 &7.30&1.0 \\
   &9&1.75 & 2.43 & 1.46 & 5.63 & 3.99 & 5.96 & 0.33 &5.97 &1.0\\
   &10&2.65 & 2.41 & 1.16 & 6.21 & 4.75 & 4.75 & 0.00 &4.76 &1.0\\
   &12&2.45 & 1.73 & 0.78 & 4.95 & 3.68 & 3.68 & 0.00 &3.68 &1.0\\
   &13&2.03 & 0.74 & 0.25 & 3.03 & 2.54 & 2.54 & 0.00 &2.55 &1.0\\
   &14&1.19 & 0.21 & 0.07 & 1.48 & 1.05 & 1.05 & 0.00 &1.03 &1.0\\
   &15&0.84 & 0.08 & 0.05 & 0.98 & 0.76 & 0.76 & 0.00 &0.76 &1.0\\
   &16&0.50 & 0.03 & 0.01 & 0.55 & 0.42 & 0.42 & 0.00 &0.43 &1.0\\
   \hline
   KBr&19&0.36 & 2.56 & 2.94 & 5.86 & 3.15 & 10.77  & 0.71 &10.97 & 0.98 \\
   &20&1.57 & 0.32 & 0.18 & 2.07 & 1.06 & 1.06 & 0.00  &1.06 & 1.0 \\
   &21&0.73 & 0.11 & 0.17 & 1.00 & 0.46 & 0.46 & 0.00  &0.46& 1.0 \\
   \hline
  \end{tabular}
 \end{center}
\end{table}

\subsubsection{Pile-up}\label{s3-3-2}
\begin{figure} [htbp]
 \begin{center}
\includegraphics[width=160mm]{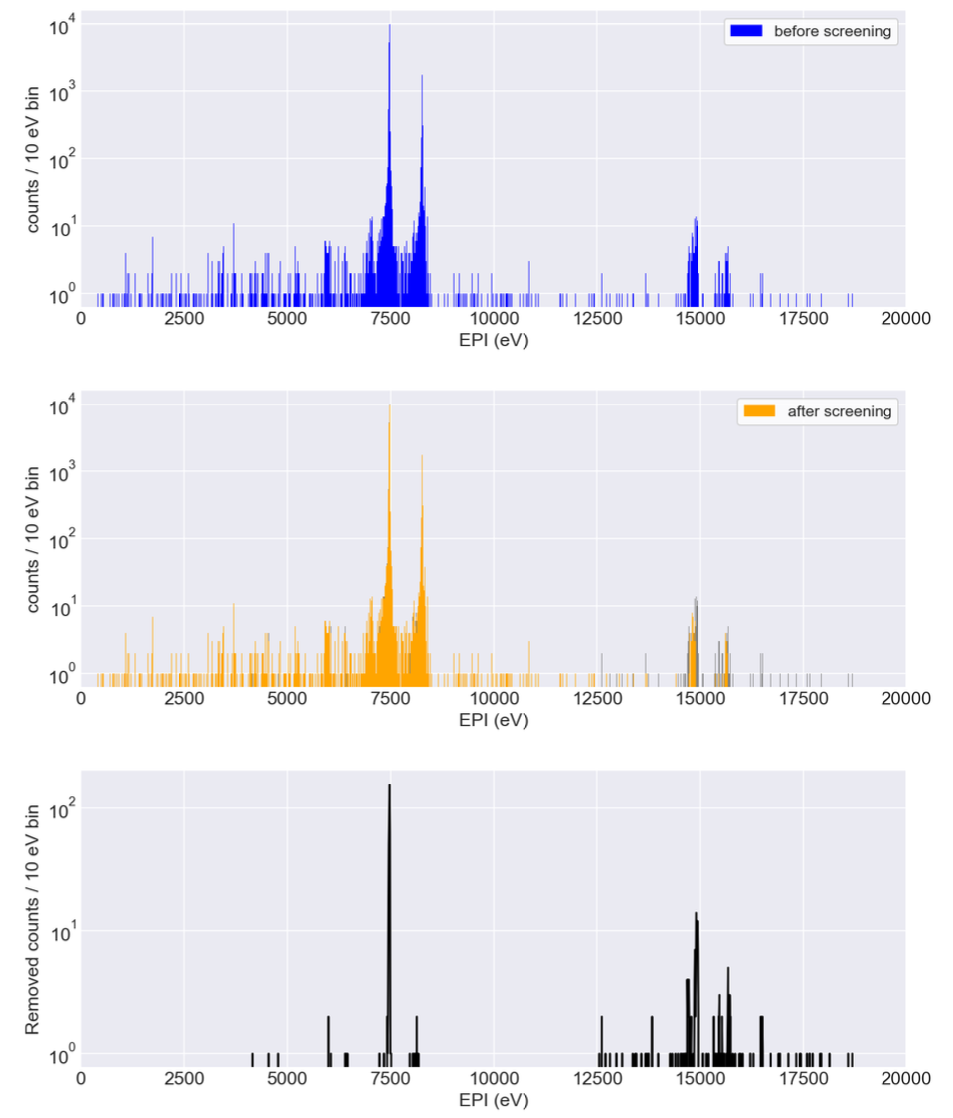}
  \caption
  { \label{fig:pileup} 
  X-ray spectra with Hp grade in step 2 in Table~\ref{tab:obs}. The peaks around
  7.5~keV and 15~keV show the energy of Ni K$\alpha$ and twice that of it, respectively. The blue line in the upper panel and the gray line in the middle panel shows the spectrum before the screening in
  Figure~\ref{fig:Rt-PH}, while the orange line in the middle panel is after screening.
  The removed counts are shown in the bottom panel.
  }    
 \end{center}
\end{figure}

When an event is detected, overlapping pulses (i.e., secondary events) are sought by subtracting the scaled average pulse shape from the primary event.
If the overlapping pulses are too close in time, they are not distinguished and treated as one event, referred to as pile-up. Similarly to CCD pile-up, this distorts the spectrum and leads to underestimation of the count rate. For an X-ray spectrum dominated by a single line, such as the one presented here, pile-up is most evident in the spectrum at an energy twice that of the line (Figure \ref{fig:pileup}).

\begin{figure} [htbp]
 \begin{center}
    \centering
    \scalebox{0.6}{\includegraphics{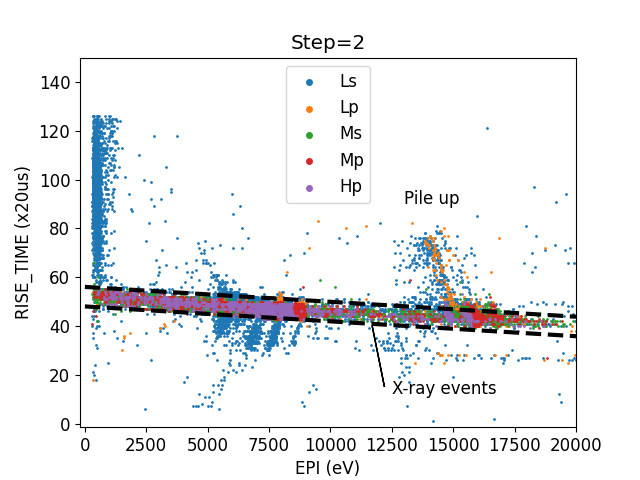}}
    \caption
    { \label{fig:Rt-PH} 
    Plots of \texttt{RISE\_TIME} versus energy of all events observed in step 2 in
    Table~\ref{tab:countrate}. Screening criteria for x-ray events are shown with
    dashed curves $|\texttt{RISE\_TIME}-52+\textrm{EPI}*(52-42)/16383.75|\leq4$, which
    removes some pile-up events seen at twice of the Ni K$\alpha$ energy.
    }    
 \end{center}
\end{figure}

For astrophysical spectra with multiple lines and continuum emission, piled-up events cannot be identified solely on the basis of energy screening. Therefore, we use the distribution of the \texttt{RISE\_TIME} versus energy relation\cite{mochizuki2024}. \texttt{RISE\_TIME} indicates how fast a pulse reaches its peak in time, which is negatively correlated with the detected photon energy. Pile-up events deviate from the normal trend (Figure~\ref{fig:Rt-PH}), allowing us to remove pile-up events. 
Furthermore, we screened events using \texttt{SLOPE\_DIFFER} in the $E<11$~keV band\cite{mochizuki2024}.
This flag indicates that the derivative is different from the averaged one, and 
is triggered under either of the following conditions: (1) the search for the secondary candidate is not completed within the expected pulse duration, or (2) the derivative exhibits a significantly negative value during the secondary pulse search.
After applying the screening, 
we also observed an improvement in the pile-up spectrum (Figure \ref{fig:pileup}), although not all pile-up events were completely removed, as seen in the overlapping distribution in Figure~\ref{fig:Rt-PH}.
Quantitatively, the screening reduced the number of events  with energies $>10$~keV to 32.8\% of the original one.

We now estimate the dead time caused by not detecting overlapping pulses too close in time. The PSP software is designed to detect two pulses separated by $>$6~ms for all and $>$2 ms when the secondary pulse is large enough to have the peak contrast to the primary pulse $>$1/20. 
Assuming that the count rate with the averaged value of $\nu$~s$^{-1}$~pixel$^{-1}$ fluctuates following the Poisson distribution, the fraction of events arriving within $\Delta t_\mathrm{thres}$ of another pulse is $1-\exp(-2 \nu \Delta t_\mathrm{thres})$. 
When we simply assume that $\Delta t_\mathrm{thres}=$2~ms,
the fraction of pile-up events, hence the loss of observing time by screening them, is 6\% for $\nu=15$~s$^{-1}$~pixel$^{-1}$. We note that $\Delta t_\mathrm{thres}$ in SCDP is as small as $\sim0.8$~ms, and this difference provides a slight discrepancy between the PSP processed rate and the SCDP rate in the highest count rate case (Table \ref{tab:countrate}).

\subsubsection{Cross Talk}\label{s3-3-3}
Untriggered electrical cross talk is considered to be the major cause of the degradation of energy resolution at very high count rates. Electrical nearest-neighbor channels cross-talk with each other, resulting in pseudo-events with small contaminating pulses (``cross-talk children'') with respect to the X-ray pulses of the parent events (``cross-talk parents'').  The impact can be to increase or decrease the inferred X-ray energy depending on the energy of the parent pulse and the displacement between pulses.  Energy resolution degrades at high rates because of the increased probability of overlap and will be worse for hard spectra than for soft spectra.
In this work, if a pulse in pixel $i$ arrives within $\pm25$~ms of another pulse in pixel $i \pm 1$ (first neighbors), this event is regarded to be contaminated by untriggered electrical cross-talk children.  This threshold can be extended to $i \pm 2$ (second neighbors).
Hereafter we call such an event as ``cross-talk-contaminated event'', and 
removing the cross-talk-contaminated events is called a ``cross-talk cut''.

\begin{figure} [htbp]
 \begin{center}
\includegraphics[width=160mm]{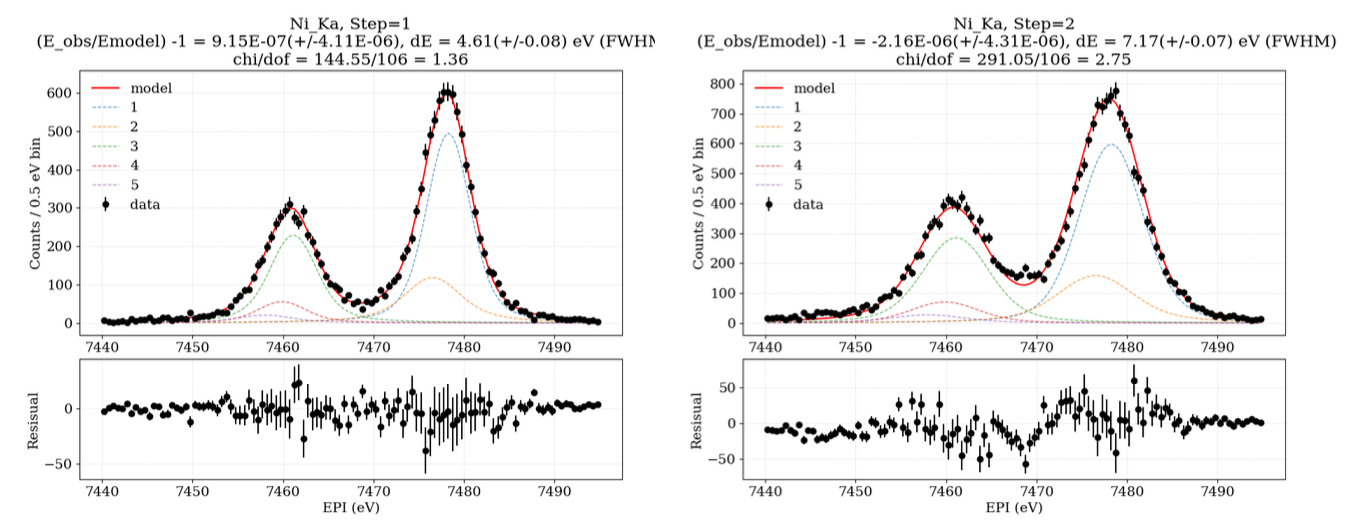}
 \end{center}
 \caption
 { \label{fig:fit_PPL} 
 Fitting result of the Ni K$\alpha$ lines in step 1 (left) and step 2 (right) in Table~\ref{tab:countrate}. The five Lorentzians (1--5) are shown with the dotted curves, while their sum is shown in the solid curve. The lower
 panel shows the residuals to the best fit.}
\end{figure}

To study this degradation, the change in energy resolution is examined. 
First, we used the uniform illumination test data.
We fit the Ni K$\alpha$ line in the 7400--7495~eV for each spectrum in Step 1 to 17. Each spectrum with each pixel and step is individually fitted and subsequently adjusted to align the center of the line with the correct energy. Then, spectra with the same step are combined to increase the signal-to-noise ratio in the spectral fitting. Figure \ref{fig:fit_PPL} shows some examples of the fitting result. The energy resolution degrades from 4.61 $\pm$ 0.06~eV with a count rate of 0.91~s$^{-1}$~pixel$^{-1}$ to 7.17 $\pm$ 0.07~eV with 14.5~s$^{-1}$~pixel$^{-1}$.

\begin{figure} [htbp]
 \begin{center}
 \includegraphics[width=160mm]{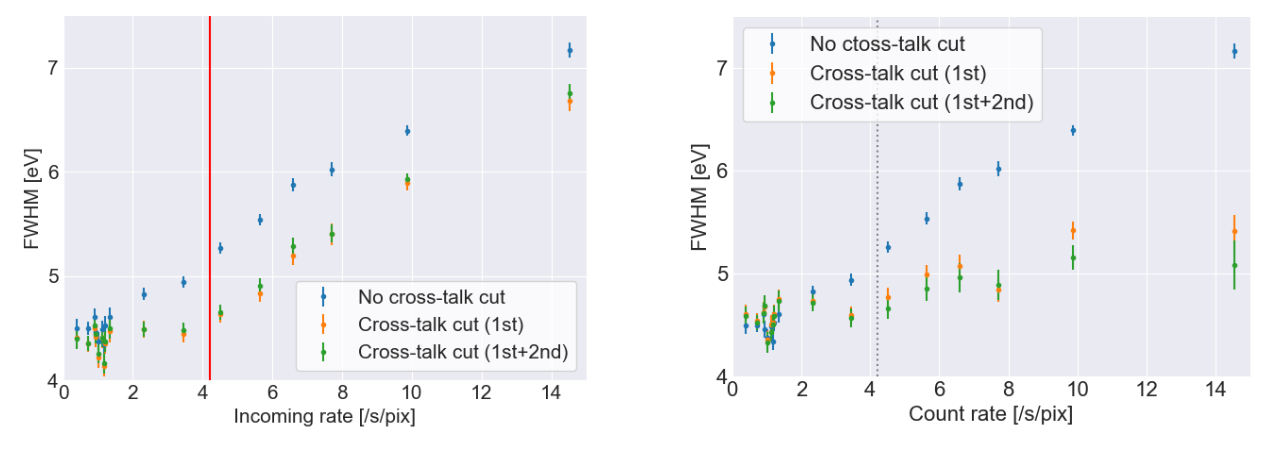}
 \end{center}
 \caption
 { \label{fig:FWHMtrend} 
 Energy resolution (FWHM) of the Ni K$\alpha$ lines as a function of the incoming rate using the PSP (left) and SCDP (right) data. The blue,
 orange, and green data respectively indicate the results without the cross-talk cut,
 with the cut for the first and the second neighbors. 
 {The red line  in the left panel show the CPU limit, which is also shown as the gray dotted line in the right panel for comparison.}}
\end{figure} 

The degradation can be somewhat recovered by applying the cross-talk cut, at a sacrifice of the effective exposure time. 
Figure \ref{fig:FWHMtrend} shows the result, where the energy resolution is plotted against the corrected count rate without the cross-talk cut (blue), with the cut for the first (orange) and second (green) neighbors.
To be accurate, the horizontal axis of Figure \ref{fig:FWHMtrend} should be the count rate of neighbor pixels. However, since the whole array is uniformly illuminated in this test, the average count rate is used instead. For the PSP result (left), the resolution monotonically degrades but is recovered by the cross-talk cut below the CPU limit.
The red vertical line shows the CPU limit line, which is 50~cts~s$^{-1}$~quadrant$^{-1}\times0.76$ (cleaned event rate) $/9$ pixel $=4.2$~s$^{-1}$~pixel$^{-1}$. The difference between the first and second neighbor cut is not significant. Beyond the CPU limit, this strategy works only partially because a some events are discarded, and information about some cross-talk parents is lost. Figure \ref{fig:ltcrv} shows that the dead time (shown in green) depends on the pixel, as the count rate varies from pixel to pixel. If the cross-talk parent is within the dead time, we can no longer identify it. The effect of the dead time is confirmed by the SCDP result, which does not suffer from such losses. Recovery of resolution is still seen beyond the CPU limit (Figure \ref{fig:FWHMtrend} right), while incomplete, presumably due to thermal bath fluctuation caused by the very strong X-ray illumination. 

\begin{figure}[htbp]
\centering
 \includegraphics[width=90mm]{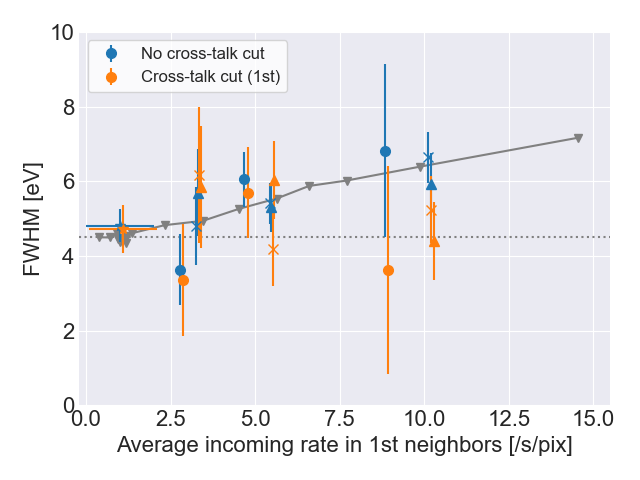}
 \caption{
 \label{fig:FWHMtrend2}
 Energy resolution (FWHM) of the Cu K$\alpha$ lines as a function of the average count rate in the 1st neighbor pixels. 
 The circle, cross, and triangle data points represent pixel 19, 20, and 21, respectively. The data points for low current conditions with a count rate of 0--2~cts~s$^{-1}$~pixel$^{-1}$ were summed to increase the signal to noise ratio.
 The data points with the cross-talk cut (first neighbor) are slightly shifted towards the positive horizontal axis for clarity.
 The gray solid line and inverse-triangle bins show the energy resolution degradation in the uniform illumination test (same as the blue bins in Figure \ref{fig:FWHMtrend}), and the gray dotted line shows FWHM=4.6~eV.
 Note that the three points on the right side of the figure ($>8$ cts/s/pix) exceed the CPU limit.}
\end{figure}

To study the cross-talk cut under the circumstance similar to actual in-orbit observations, the data with the point-like illumination are also examined.
All electrically contiguous pixels must have sufficient counts for this study. We found that only five pixels, from pixel 18 to 22, met this criterion in Steps 31--37. For example, in pixel 19, the untriggered cross talk children from pixel 18 or 20 can contaminate the event pulse.
We created spectra in each step and each pixel and fitted the Cu K$\alpha$ model to evaluate the energy resolution, as was done in the uniform illumination test data. The fitting results are shown in Figure \ref{fig:FWHMtrend2}. We focus solely on the 1st neighbor cross-talk cut. For comparison, we overlay the degradation of resolution obtained in the uniform illumination test by the gray solid line. The PSP overflow occurs in the most right-handed data bins.
Despite large error bars due to poor statistics, {the resolution trend as a function of the count rate in the neighbor pixels is consistent with Figure \ref{fig:FWHMtrend} even when simulating a point source, and the resolution appears to improve with the cross-talk cut, approaching the nominal performance (4.6~eV).}

%%%%%%%%%%%%%%%%%%%%%%%%%%%%%%%%%%%%%%%%%%%%%%%%%%%%%%%%%%%%%%%%%%%%%%
\section{Simulation}\label{s4}
%%%%%%%%%%%%%%%%%%%%%%%%%%%%%%%%%%%%%%%%%%%%%%%%%%%%%%%%%%%%%%%%%%%%%%
In this section, we construct models to describe the high count rate effects (\S~\ref{s4-1}). The models are intended for observation planning, thus are phenomenological. In combination with the event simulator in the \texttt{HEAsoft} package, observers can assess the level of high count rate effects with the presented models, as exemplified in \S~\ref{s4-2}. 

\subsection{Models describing high count rate effects}\label{s4-1}
In this subsection, we simulate a point source observation. It is assumed that the mirror has the 1.3 arcmin HPD\cite{Okajima2016}, the source is positioned on one of the four central pixels of the \textit{Resolve} array, and the roll angle is 0.

\subsubsection{CPU limit}\label{s4-1-1}
We model the CPU consumption rate of PSP. We assume that the rate is proportional to the incoming count rate of each grade (HR, MR, and LR) plus the base load. The four CPUs (A0, A1, B0, and B1) do not have identical loads: A0 and B0 handle communication with the Xbox, while A1 and B1 process the anti-co data. A1 receives counts from PIXEL 12, which produces $\sim$6.1~s$^{-1}$~pixel$^{-1}$ of counts from the $^{55}$Fe calibration source at this experiment, in 2022. Additionally, all CPUs process baseline events, in which optimal filtering is applied to noise records for diagnostic purposes.

\begin{figure} [htbp]
 \begin{center}
\includegraphics[width=160mm]{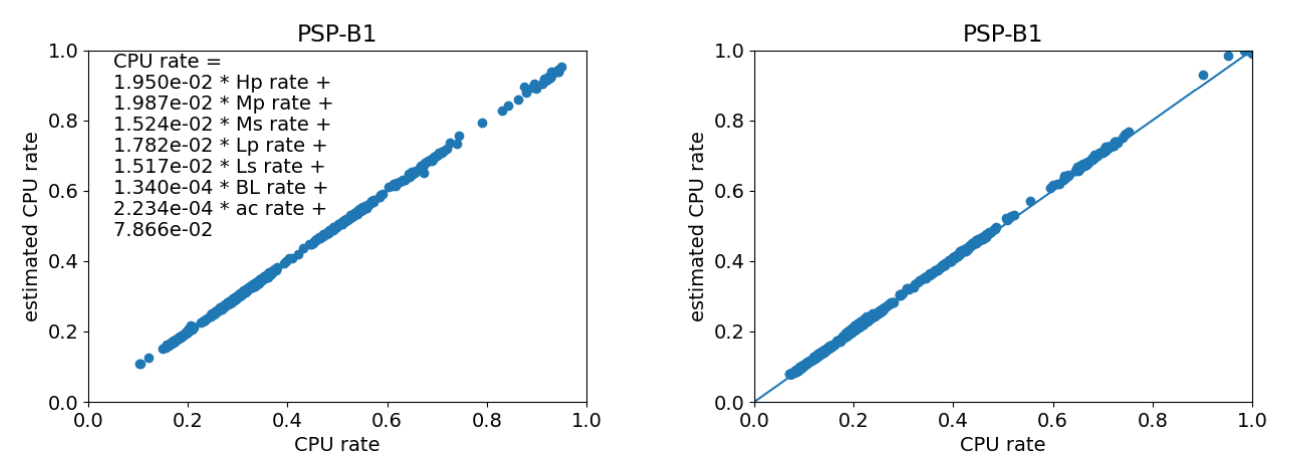}
 \end{center}
 \caption
 {\label{fig:CPUrate} 
 Result of CPU consumption model using the Ni data (left) and the application of the
 model to the KBr data (right). The result for the B1 CPU is shown. The measured CPU
 rates are plotted against the CPU rate calculated by the model. If the model works well, the data points in the right panel follows the $y=x$ line.}
\end{figure}

The CPU consumption rate is modeled as a linear combination of these rates \cite{ishisaki18a}.
\begin{equation}
\textrm{(CPU rate)}^{\mathrm{(i)}} = \sum_{j\in i, k} \, a_k^{\mathrm{(i)}} p_\mathrm{j,k} + c^{\mathrm{(i)}}, \label{eq:cpurate}
\end{equation}
where $i$ is the PSP quadrant (A0, A1, B0, and B1), $j$ is the pixel number including in each $i$, $k$ is the event grade (Hp, Mp, Ms, Lp, Ls, baseline, and anti-co events), $p_\mathrm{j,k}$ is the count rate for each pixel and event grade, 
and $c$ is   the base load. 
All of this information is available from the PSP telemetry data. We used the Ni data with CPU consumption rates ranging from 5\% to 95\% (Step 1, 3, 5, 8, and 11--17) {to estimate the coefficients by the least squares method. The results are listed in Table~\ref{tab:CPUrate_coeff}. The best-fit model based on Equation (\ref{eq:cpurate}) and the derived coefficients} matches well with observations (Figure~\ref{fig:CPUrate} left panel). For verification, we applied the model to all of the KBr data from Steps 18 to 21 (Figure~\ref{fig:CPUrate} right panel), and observed a good match as well. The result shows that the base load ($c$) ranges from 7\% to 10\%, and each x-ray event (from Hp to Ls) consumes 1.4 to 2.3\%, depending on the grade. When the anti-co rate is extremely large, for example, in the South Atlantic Anomaly, the CPU consumption can be overwhelmed by it.

\begin{table}[htbp]
 \begin{center}       
  \caption{Fitting results of the CPU load fraction} 
  \label{tab:CPUrate_coeff}
  \begin{tabular}{lcccc} 
   \hline
   \hline
   CPU & A0 & A1 & B0 & B1 \\
   \hline
   Pixel & 0--8 & 9--17 & 18--26 & 27--35\\
   \hline
   $a_\textrm{Hp}$      & $0.0197\pm0.0001$ & $0.0193\pm0.0001$ & $0.0199\pm0.0001$ & $0.0195\pm0.0001$ \\
   $a_\textrm{Mp}$      & $0.0180\pm0.0006$ & $0.0166\pm0.0005$ & $0.0194\pm0.0004$ & $0.0199\pm0.0005$ \\
   $a_\textrm{Ms}$      & $0.0174\pm0.0005$ & $0.0178\pm0.0004$ & $0.0170\pm0.0005$ & $0.0153\pm0.0003$ \\
   $a_\textrm{Lp}$      & $0.0221\pm0.0006$ & $0.0168\pm0.0006$ & $0.0174\pm0.0006$ & $0.0178\pm0.0005$ \\
   $a_\textrm{Ls}$      & $0.0147\pm0.0003$ & $0.0157\pm0.0003$ & $0.0163\pm0.0003$ & $0.0152\pm0.0002$ \\
   $a_\textrm{BL}$      & $0.0002\pm0.0001$ & $0.0000^{+0.0001}$ & $0.0001\pm0.0001$ & $0.0001\pm0.0001$ \\
   $a_\textrm{antico}$  & ---    & $0.0005\pm0.0001$ &  ---   & $0.0002\pm0.0001$ \\
   $c$                  & $0.0842\pm0.0003$ & $0.0972\pm0.0005$ & $0.0834\pm0.0003$ & $0.0787\pm0.0004$ \\
   \hline
  \end{tabular}
 \end{center}
\end{table} 

\subsubsection{Pile-up}\label{s4-1-2}
\begin{figure}[th]
\centering
\includegraphics[width=90mm]{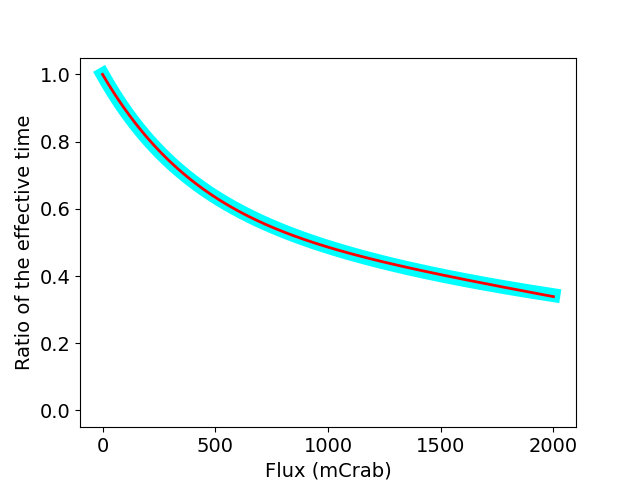}
  \caption{
 \label{fig:pileup_time}
 Ratio of the effective exposure time reduced by the pile-up effect as a function of the x-ray flux. The cyan shows
 the simulation results, whereas the red is the phenomenological fitting.}
\end{figure}

The effective time reduced by the pile-up effect on
the entire array is calculated as a function of flux from a point source.  
Once the source flux is given, the incoming rate in each pixel is derived and the effective time ratio
of $\exp(-2 \nu \Delta t_\mathrm{thres})$ (see \S~\ref{s3-3-2}) is calculated. For simplicity $\Delta t_\mathrm{thres}$ is assumed to be 2~ms at any case. Finally, the total effective time weighted by the incoming rate is calculated.
Figure \ref{fig:pileup_time} shows the ratio of the
effective exposure time ($\alpha_\mathrm{noPileUp}$) as a function of the x-ray flux
($f_\mathrm{X}$ [mCrab]) in the gate-valve-open case.  The phenomenological fitting results is
$\alpha_\mathrm{noPileUp} =
\exp(-f_\mathrm{X}/959) * 
(1-1.055\times10^{-4}f_\mathrm{X} +
4.842\times10^{-7}{f_\mathrm{X}}^2)$. Note that this relation is valid only up to
2~Crab.

\subsubsection{Cross talk}\label{s4-1-3}
We aim to construct a phenomenological model to describe the energy resolution degradation due to the untriggered cross-talk children, using the Ni data set.

First, the excess broadening due to the cross-talk effect (FWHM) is calculated to be  
\begin{equation}
 \mathrm{FWHM_{excess}}=(\mathrm{FWHM_{withXtalk}}^2-\mathrm{FWHM_{noXtalk}}^2)^{1/2},
\end{equation}
where $\mathrm{FWHM_{withXtalk}}$ is the observed energy resolution before the cross-talk cut (the blue bins in the left panel of Figure \ref{fig:FWHMtrend}) and $\mathrm{FWHM_{noXtalk}}$ is the one after the cross-talk cut (the orange bins in the right panel of Figure \ref{fig:FWHMtrend}).
Both are measurable from the ground test, and thus we can calculate $ \mathrm{FWHM_{excess}}$ for each pixel. We evaluate the energy resolution degradation up to 1st neighbor pixels, because there is little difference  between 1st neighbor cross-talk cut and 1st+2nd one (Figure \ref{fig:FWHMtrend}).

\begin{figure} [htbp]
 \begin{center}
  \begin{tabular}{cc} 
   \begin{minipage}[t]{80mm}
    \centering
    \scalebox{0.5}{\includegraphics{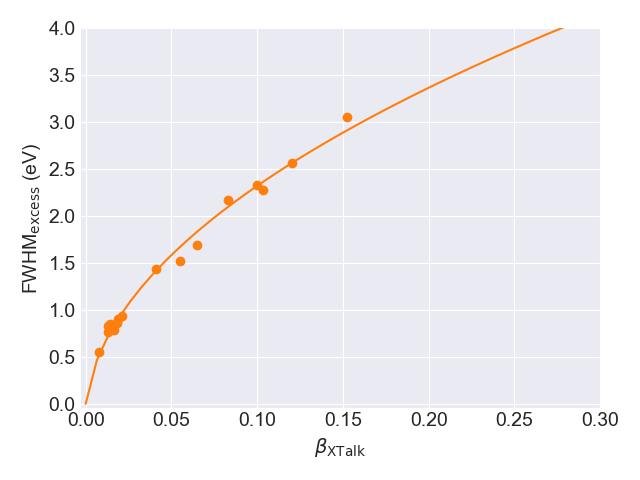}}
    \caption
    {   \label{fig:dE_CTratio}
  Relation between $\mathrm{FWHM_{excess}}$ and $\beta_\mathrm{XTalk}$ for each pixel. 
The solid lines show the phenomenological model with the best-fitting parameters (equation \ref{eq:eq2}). Note that this plot is built using the Ni data set.
   }    
   \end{minipage} 
   &
   \begin{minipage}[t]{80mm}
    \centering
    \scalebox{0.5}{   \includegraphics{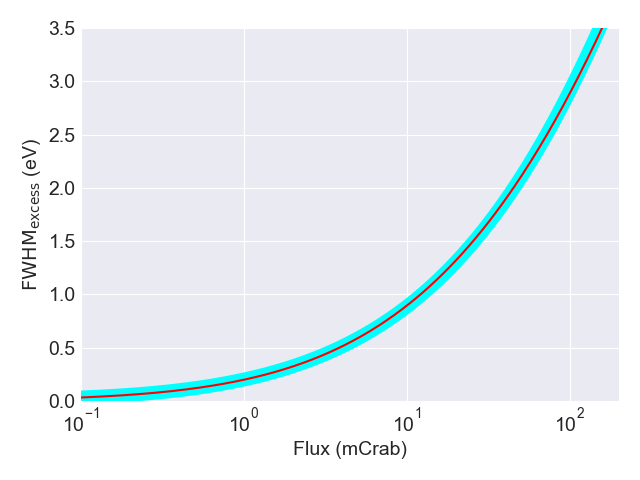}}
  \caption{
 \label{fig:Xtime_flux}
 The  averaged $\mathrm{FWHM_{excess}}$ weighted by count rates as a function of the X-ray flux (cyan). The red curve is a phenomenological function to describe the cyan curve. Note that this plot is built using the Ni data set.}
   \end{minipage} \\ 
  \end{tabular}
 \end{center}
\end{figure}

It is expected that $\mathrm{FWHM_{excess}}$  depends on the the ratio of cross-talk-contaminated events to the total ones ($\beta_\mathrm{Xtalk}$), so that we attempt to build a phenomenological model of $\mathrm{FWHM_{excess}}$ as a function of $\beta_\mathrm{Xtalk}$.
We also note that $\beta_\mathrm{Xtalk}$ is measurable because we can identify all cross-talk parents using the SCDP data.
Figure \ref{fig:dE_CTratio} shows their relation, which can be well described as
\begin{equation}
0.125*\mathrm{FWHM_{excess}}^2+0.054*\mathrm{FWHM_{excess}} = \beta_{\mathrm{XTalk}} \label{eq:eq2}
\end{equation}
with the unit of eV for $\mathrm{FWHM_{excess}}$.

The $\beta_\mathrm{XTalk}$ value in each pixel is determined from the count rate of the neighbor pixels. 
When the X-ray flux of the point source is given,
we can get a count rate and we can estimate $\beta_\mathrm{XTalk}$ for each pixel.
By substituting $\beta_\mathrm{XTalk}$ into equation (\ref{eq:eq2}), the averaged $\mathrm{FWHM_{excess}}$ weighted by the count rate is derived.
The cyan line in Figure \ref{fig:Xtime_flux} shows the result.
We have found that this curve is well described as
\begin{equation}
\log_{10}(\mathrm{FWHM_{excess}})= -0.0728 * (\log_{10} f_\mathrm{X})^2 + 0.7280 * (\log_{10} f_\mathrm{X}) -0.7041
\end{equation}
with the unit of eV and mCrab for $\mathrm{FWHM_{excess}}$ and $f_\mathrm{X}$, respectively.

\begin{figure}[htbp]
\centering
\includegraphics[width=12cm]{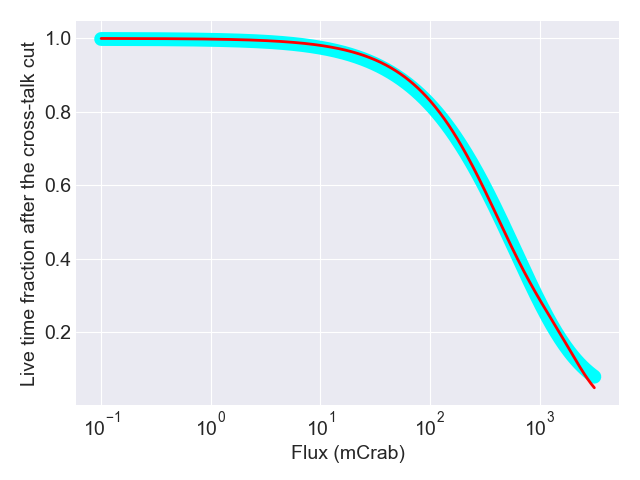}
    \caption
    { \label{fig:XT_eff} 
  Ratio of the effective exposure time ($\alpha_\mathrm{XTalk}=1-\beta_{\mathrm{XTalk}}$) as a function of the x-ray flux. The cyan shows
 the simulation results and the red is the phenomenological fitting. The loss of the effective time due to the pile-up is not included. 
   }  
\end{figure}

If cross-talk cut is performed, while the energy resolution is restored, the effective exposure time is reduced.
Now we define the live-time fraction after the cross-talk cut as $\alpha_\mathrm{XTalk}=1-\beta_{\mathrm{XTalk}}$.
Since $\beta_{\mathrm{XTalk}}$ for each pixel and the count rate map from a point source can be calculated for a given X-ray flux,
the averaged $\alpha_\mathrm{XTalk}$ weighted by count rates is estimated as a function of the X-ray flux.
Figure \ref{fig:XT_eff} shows the result. The phenomenological fitting results in
\begin{equation}
\alpha_\mathrm{XTalk} = \exp(-f_\mathrm{X}/613) * (1-2.358\times10^{-4}f_\mathrm{X} + 6.919\times10^{-7}{f_\mathrm{X}}^2). \label{eq:alpha}
\end{equation}

Equations (\ref{eq:eq2}) and (\ref{eq:alpha}) provide a user with information to decide whether to perform the cross-talk cut or not.
For example, when we observe a point source whose flux is 400~mCrab with the 1/4 ND filter (that is, when the irradiated flux is 100~mCrab), the line distortion with FWHM of 2.9~eV is produced without the cross-talk cut. On the other hand, if the cross-talk cut is performed, the live-time fraction becomes 83.5\%.

It is noteworthy that this degradation is highly dependent on spectral hardness. In the Ni data set,  its continuum level is much fainter and the Ni lines are much stronger than the actual astronomical data, resulting in a very hard spectral shape. When we estimate the energy resolution distortion for astronomical observations, the $\mathrm{FWHM_{excess}}$ has to be adjusted by the photon count-weighted average energy, as exemplified in \S\ref{s4-2-2}.
It should be noted that the energy resolutions vary across the pixels. Users can extract a spectrum from each pixel and see the energy resolution change. 

\subsection{Simulated astrophysical observation}\label{s4-2}
\subsubsection{Input spectrum}\label{s4-2-1}
We use GX 13$+$1 as a test case. The source is a LMXB hosting a neutron star (NS) and a
low-mass companion with an orbital period of 24.5~days\cite{iaria2014}. The X-ray flux
fluctuates around $\sim0.3$~Crab. During periodic dips, which may be related to a super-orbital periodicity, the soft band flux is reduced by $\sim$4 times
\cite{iaria2014}. The spectrum exhibits absorption by the ionized Fe and Ni K
lines\cite{sidoli2002}.

\begin{figure} [htbp]
 \begin{center}
  \begin{tabular}{cc} 
   \begin{minipage}{80mm}
    \centering
    \scalebox{0.3}{\includegraphics{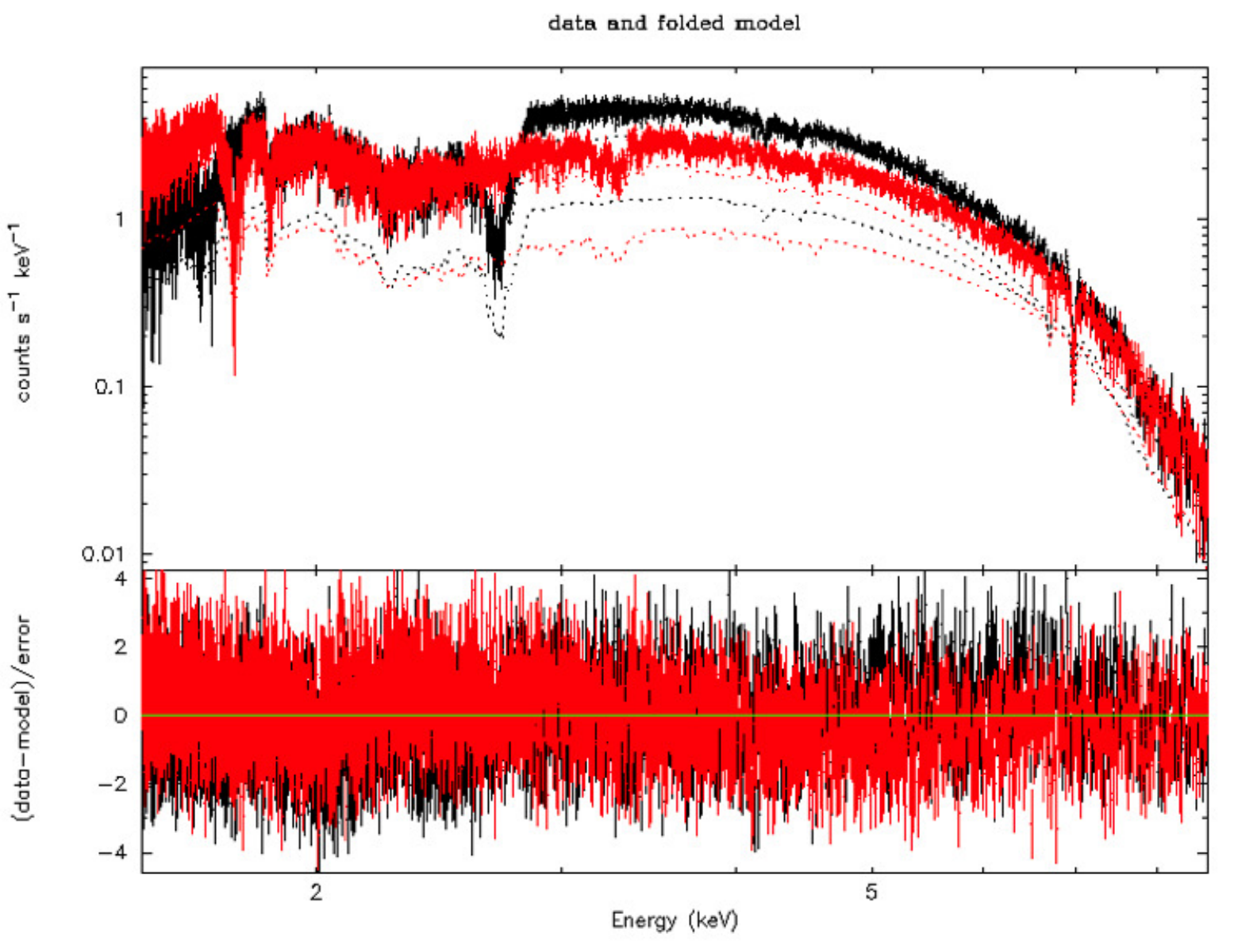}}
    \caption
    {\label{fig:GX13_HETG} 
    GX 13$+$1 spectrum observed with \textit{Chandra} HETG and its best-fit model in the
    upper panel and the residuals to the fit in the lower panel. The black and red
    colors are for the $\pm$1st order dispersion data.
   }
   \end{minipage} 
   &
   \begin{minipage}{80mm}
    \centering
    \scalebox{0.5}{\includegraphics{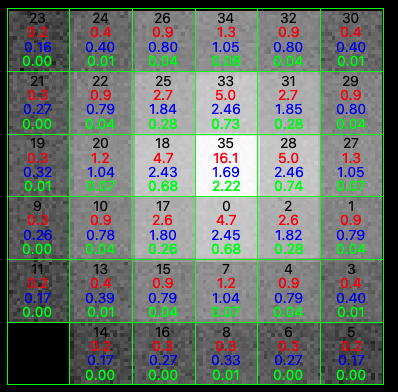}}
    \caption
    {\label{fig:heasim_image} 
    Distribution of count rate of the simulated \textit{Resolve} observation of GX 13+1. The pixel
    number is given in black, while the total, Hp, and Mp count rates
    (s$^{-1}$~pixel$^{-1}$) are in red, blue, and green texts, respectively.}
    \end{minipage} \\ 
  \end{tabular}
 \end{center}
\end{figure}

We start from the \textit{Chandra}  High-Energy Transmission Grating (HETG) observation with the Advanced CCD Imaging Spectrometer (ACIS) operated in the
continuous clocking mode (OBSID: 11818). The x-ray flux was 270~mCrab. We extracted the
X-ray spectra and fitted them (Figure~\ref{fig:GX13_HETG}) using two continua (disk blackbody for the accretion disk and blackbody blackbody for the NS surface emission) attenuated
by Galactic absorption and local absorption by the ionized outflow. The H- and
He-like Fe K absorption features are described using the \texttt{kabs}
model\cite{tomaru2020}. The two absorption features are fitted with the absorption
column of $8.9\times10^{18}$ and $2.5\times10^{17}$~cm$^{-2}$ (for each ion) with a
common turbulent velocity of 200~km~s$^{-1}$ and an outflow velocity $-330$~km~s$^{-1}$. Constraining the absorption features is an important science goal of
\textit{Resolve}.

The mock event file was made using the \texttt{heasim}  event simulator for a 30 ks exposure with the source placed on one of the center pixels (pixel 35).  The best-fit spectral model for the Chandra HETG observation was used as a spectrum input, with the flux reduced by a factor of four, assuming the use of the ND filter equipped with \textit{Resolve}.
Events whose energies exceed 130.5 eV (mean energy corresponding to a derivative-trigger threshold of 120) are extracted. The count rates for each grade in each
pixel are calculated by \texttt{rslbranch}, which are shown in
Figure~\ref{fig:heasim_image}. Over the entire array, the total, Hp, and Mp count rates
are 62.5, 33.3, and 6.8 s$^{-1}$~array$^{-1}$, respectively. The branching ratio of the
Hp events is 0.53. For each pixel, the total count rate varies by nearly two orders of magnitude from 16.1 (on-source
position) to 0.2~s$^{-1}$~pixel$^{-1}$ (outer pixels). 

\subsubsection{Estimates of high count rate effects}\label{s4-2-2}
\paragraph{CPU limit}
The total count rate in the upper right quadrant (PSP-B1) is
33.46~s$^{-1}$~quadrant$^{-1}$, which is the highest because the quadrant includes the most illuminated
pixel 35. The CPU consumption rate is calculated using Equation (\ref{eq:cpurate}).  Here, the
anti-co and baseline counts are ignored (see \S~\ref{s4-1-1}).  The calculated CPU load
is 0.306 (PSP-A0), 0.206 (A1), 0.307 (B0), and 0.667 (B1). Therefore, no event loss is expected in this observation.

\paragraph{Pile-up}
In actual observations, the piled-up events are merged into a single event, but in the simulator, each is detected separately and flagged.
The flagged events are listed in the \texttt{heasim} output file. We set the parameter
of \texttt{dtpileup} as 0.002, which means that we simply assume that
$\Delta t_\mathrm{thres}=$2~ms in any case.
The pile-up fraction, thus the loss of the
effective exposure time, is calculated to be 2.5\%.

\paragraph{Cross talk}
The \texttt{heasim} tool does not simulate the degradation in energy resolution caused by the cross talk. Hence, we must estimate it using the empirical equation developed in \S\ref{s4-1-3}. It is important to note that this estimation is based on the Ni data set and that the degradation is heavily influenced by the spectral hardness.
For a preliminary approximation, we scaled $\mathrm{FWHM_{excess}}$ by the ratio of the average energy of detected counts in GX 13+1 ($E_\mathrm{GX13+1}=4.14$ keV) to that in the Ni dataset ($E_\mathrm{Ni\mathchar`-K}=7.47$ keV).

\begin{figure} [htbp]
 \begin{center}
  \begin{tabular}{cc} 
   \begin{minipage}[t]{80mm}
    \centering
    \scalebox{0.55}{\includegraphics{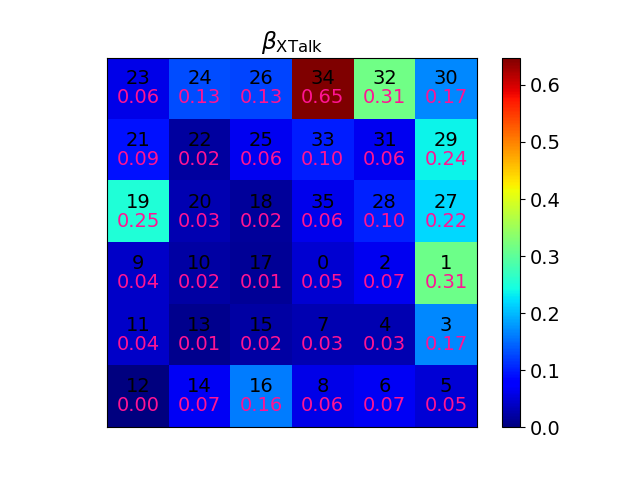}}
    \caption
    { \label{fig:CTratiomap} 
    Pixel map of the ratio of the cross-talk contaminated events: $\beta_{\mathrm{XTalk}}$
    in Equation~\ref{eq:eq2}. The black shows the pixel number. 
   }    
   \end{minipage} 
   &
   \begin{minipage}[t]{80mm}
    \centering
    \scalebox{0.55}{\includegraphics{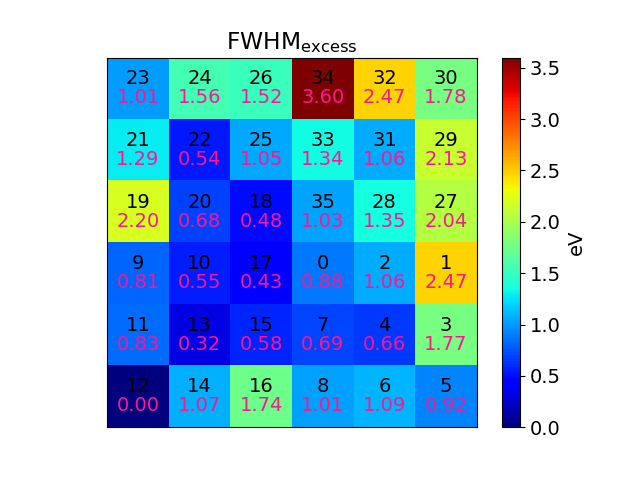}}
    \label{fig:CT_FWHMmap} 
    \caption
    { 
    Pixel map of the excess broadening: $\mathrm{FWHM_{excess}}$ calculated with Equation~\ref{eq:eq2} multiplied by a correction factor of  $E_\mathrm{GX13+1}/E_\mathrm{Ni\mathchar`-K}=0.58$.
    The black shows the pixel number.
     Note that the value is extrapolated if $\beta_{\mathrm{XTalk}}>0.20$.
   }    
   \end{minipage} \\ 
  \end{tabular}
 \end{center}
\end{figure}

The excess broadening, $\mathrm{FWHM_{excess}}$, of a pixel increases when neighboring pixels exhibit high count rates. In the current simulation, the excess broadening is most pronounced in pixel 34, which is the first neighbor of the most illuminated pixel, 35. This effect must be assessed pixel by pixel. 
{Initially, we determine the ratio of cross-talk-contaminated events ($\beta_{\mathrm{XTalk}}$ in Equation~\ref{eq:eq2}) based on the simulated event file. Subsequently, utilizing Equation~(\ref{eq:eq2}), we estimate the extent of resolution degradation for each pixel.} Here, we adjust $\mathrm{FWHM_{excess}}$ by a factor of $E_\mathrm{GX13+1}/E_\mathrm{Ni\mathchar`-K}=0.58$. The outcomes are depicted in Figures~\ref{fig:CTratiomap} and \ref{fig:CT_FWHMmap}. Finally, {we compute the average degradation of the energy resolution  weighted by the Hp count rate}. The resulting $\mathrm{FWHM_{excess}}$ for the entire spectrum is estimated at 1.19 eV. Considering a 5 eV resolution, the degraded resolution is estimated to be 5.14 eV.

\subsubsection{Astrophysical impact}\label{s4-2-3}
\begin{figure} [htbp]
 \begin{center}
  \begin{tabular}{ccc} 
   \begin{minipage}[t]{80mm}
    \centering
    \scalebox{0.37}{\includegraphics[angle=270]{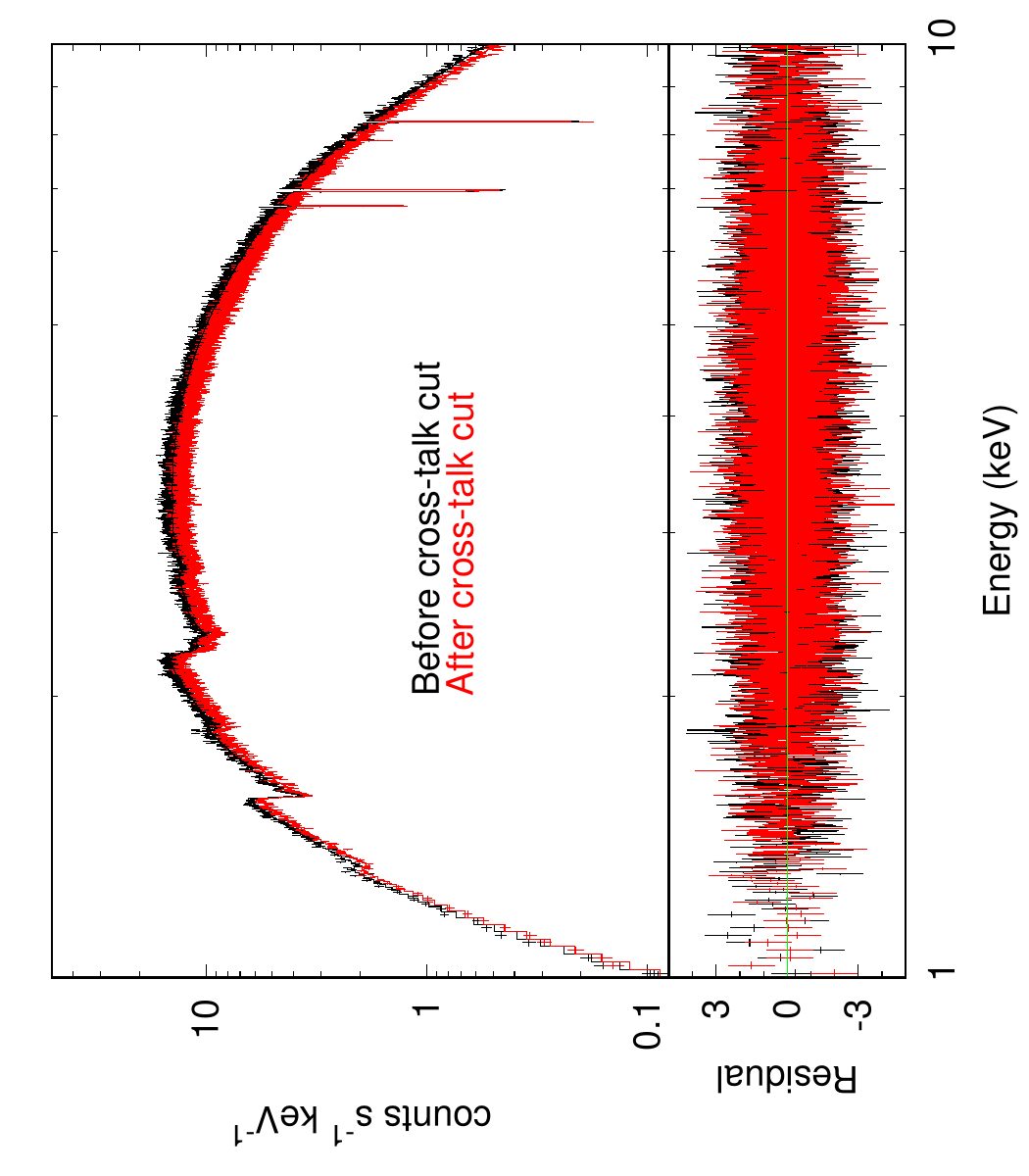}}
    \caption
    { 
    \label{fig:spec2} 
    The simulated GX 13+1 spectra with Hp grade, based on the 5~eV energy resolution
    response. The black and red are the spectra before and after the cross-talk cut,
    respectively. The lower panel shows the residuals between the simulated data and the model.  We adopt the total duration time as the exposure time. {The spectra are rebinned for easier viewing.}
   }   
   \end{minipage} 
   &
   \begin{minipage}[t]{80mm}
    \centering
    \scalebox{0.37}{\includegraphics[angle=270]{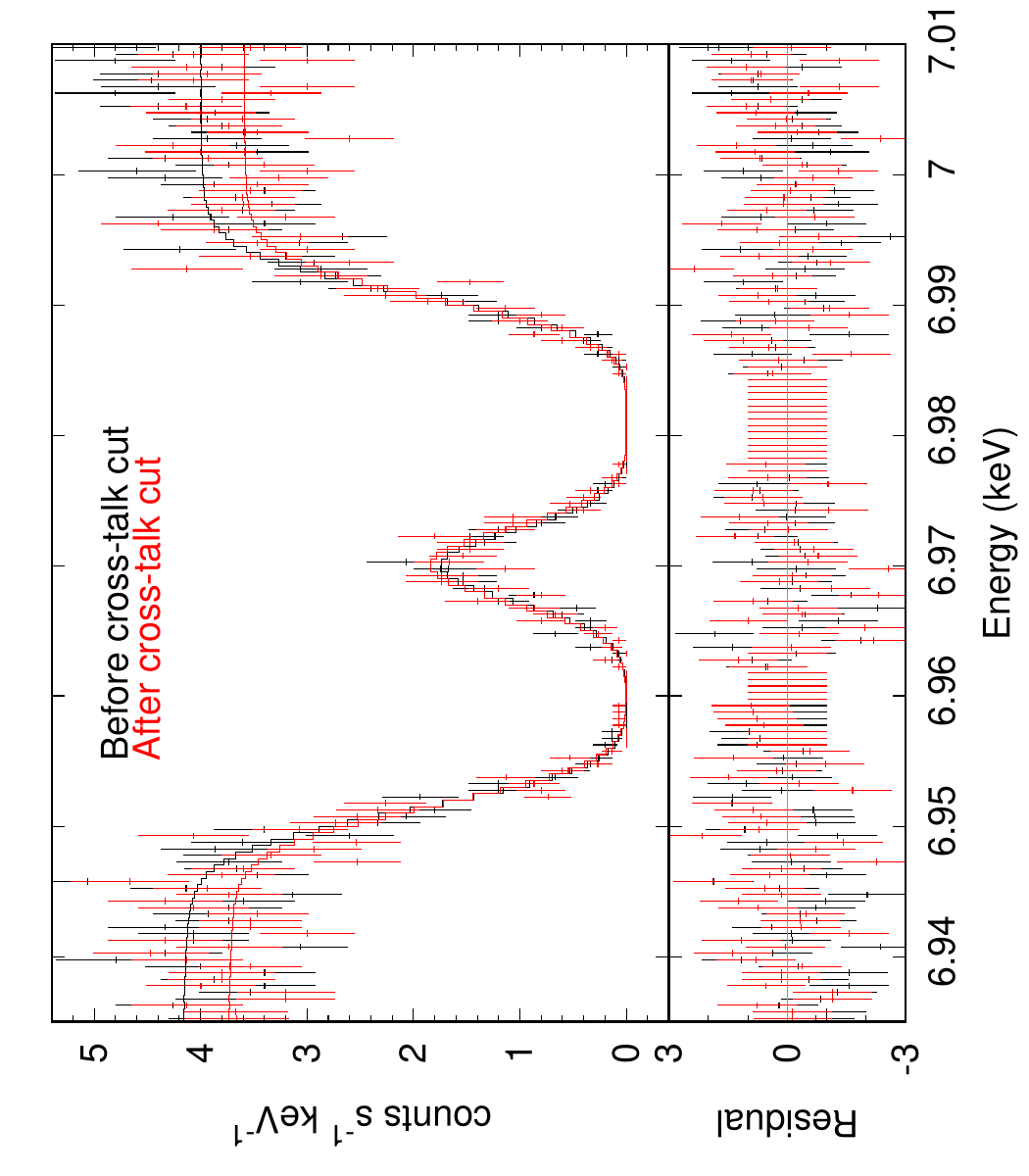}}
    \caption
    { 
    \label{fig:spec} 
    Same as Figure \ref{fig:spec2}, but only the energy band around the H-like Fe-K
    absorption lines are shown.
    The lower panel shows the data minus the folded model.
    The fitting result for the red spectrum shows $v_\mathrm{turb}=200$~km~s$^{-1}$ as expected, but the one for the black results in $v_\mathrm{turb}=206$~km~s$^{-1}$, if the spectral distortion caused by cross talk is not corrected.
    }
   \end{minipage} \\ 
  \end{tabular}
 \end{center}
\end{figure}

We have visualized how cross talk affects observations. Figures \ref{fig:spec2} and \ref{fig:spec} display the simulated spectra of GX 13+1, obtained with a 30 ks exposure time at a flux of 270 mCrab and using the 1/4 ND filter. To flexiblely adjust the energy resolution, spectra were generated using the \texttt{fakeit} command in \texttt{xspec}, based on the best fit model of the \textit{Chandra}/HETG data (depicted in Figure \ref{fig:GX13_HETG}). The black curve represents the spectrum without the cross-talk cut. The response function with a resolution of 5 eV was used and the input model was artificially smoothed using the \texttt{gsmooth} model to explain the line distortion. In contrast, the red curve illustrates the spectrum after the cross-talk cut. While the model spectrum flux was reduced, the energy resolution is restored.

When the turbulent velocity of the absorption
line is 200~km~s$^{-1}$, and if the fitting is performed without recognizing the 1.19~eV
excess broadening by the cross talk, the derived turbulent velocity becomes $\left(
\left(200\mathrm{\,km\,s}^{-1}\right)^2 + (1.19\mathrm{\,eV} /6970\mathrm{\,eV} \times
c)^2\right)^{1/2}=206$~km~s$^{-1}$, where $c$ is the light speed. If the cross-talk cut is implemented, the energy resolution will remain 5~eV, while the Hp rate decreases from 33.3~s$^{-1}$~array$^{-1}$ to 29.9~s$^{-1}$~array$^{-1}$, resulting in the live time fraction of 89.8\%. The trade-off between these factors should be considered for each science case using the methodology outlined here.

The same simulation was repeated with different fluxes. If the flux is doubled, the estimated CPU consumption rate is 1.21 on PSP-A1. In this case, the situation worsens. Event loss occurs and the number of events that can be processed is reduced by a factor of 1/1.21. In addition, some cross-talk parents cannot be identified, and the cross-talk cut cannot be completely applied. The total, Hp, and Mp processed rates are 96.9, 40.1, and 15.0 s$^{-1}$ array$^{-1}$, respectively. The average $\mathrm{FWHM_{excess}}$ weighted by the Hp count rate is calculated to be 1.59 eV, and the derived turbulent velocity becomes $211$ km s$^{-1}$. If the flux is tripled, the estimated CPU rate is 1.74 on PSP-A1. The total, Hp, and Mp incoming rates in the array are 145.3, 45.8, and 16.0 s$^{-1}$ array$^{-1}$, respectively. The average $\mathrm{FWHM_{excess}}$ weighted by the Hp count rate is 1.75 eV, and the turbulent velocity derived becomes $214$ km s$^{-1}$ (see Figure \ref{f01}).

In studies of x-ray binaries, including GX 13+1, spectra at different flux levels
are often compared for their physical interpretation. If the effect of the cross talk is
not considered properly, it may lead to a wrong conclusion, such as that the absorption
lines are broader (i.e., the turbulent velocity is larger) when the spectrum is
brighter. When performing such an analysis, the
cross-talk cut must be performed properly.  
When analyzing very bright
objects that may exceed the CPU limit (e.g., Cyg X-1), it is not possible to identify
all the cross talk events. 
{One of the  possible solutions is to calculate the degradation of energy resolution and evaluate whether the observed line broadening is significantly larger than it. If so, the observed line broadening would indicate the intrinsic changes in the target.}

\appendix

\section{Event loss type and event buffer in PSP}\label{sec:PSPbuffer}

This section details the event processing flow within the CPU in the PSP, explains the function of the event buffer, and discusses the locations where different types of event losses, characterized by \texttt{EL\_REASON}, occur.

The signals digitized on the Xbox are inputted into the FPGA in PSP, where the information from these digital signals is stored in two distinct buffers for each pixel. The first buffer is the Wave Form Ring Buffer (WFRB), which continuously stores both the ADC samples and their time derivatives as calculated by the FPGA, independent of event triggers. The WFRB has a recording capacity of 20.97 seconds. Once this buffer is full, the write position returns to 0, and the incoming samples overwrite the existing data.

The second buffer is the First-In First-Out buffer (FIFO). Within the FPGA, event candidates are identified from the time-series data, and trigger information is saved in the Event Dual Buffer (EDB) located in the FPGA. The CPU reads this trigger information from the EDB and transfers the corresponding ADC samples and their time derivative from the WFRB to the FIFO. Specifically, the ADC sample at the trigger point, along with the 219 samples preceding it and the 874 samples following it, is combined to form a single waveform dataset of 1024 ADC samples, which is stored in the FIFO. The FIFO can hold up to 256 event candidates.

If the CPU is unable to process events quickly enough, and the FIFO is full when a new event arrives, all 257 event candidates (including the new one and the 256 already accumulated) are discarded. This scenario is classified as an event loss with \texttt{EL\_REASON}=1. Alternatively, event loss can occur if the data in the WFRB have  been overwritten before the CPU retrieves it, which is classified as \texttt{EL\_REASON}=2.

In the case of \texttt{EL\_REASON}=1, it is possible to determine that 257 events are lost each time it occurs. However, for \texttt{EL\_REASON}=2, the number of discarded events cannot be quantified because it depends on the count rate. To address this, the start and end times of these event losses are recorded as bad time intervals during data processing, which is used to correct the exposure time.

Furthermore, if the EDB becomes full due to FPGA-triggered event candidates, this corresponds to \texttt{EL\_REASON}=0. However, such an occurrence would require an average count rate of $\gtrsim 111$~cts~s$^{-1}$~pix$^{-1}$ in average\cite{ishisaki18a}, which is not expected during actual observations. Therefore, this case is not considered in this paper. It should also be noted that this situation did not arise during the ground experiments conducted.

% \disclosures 
\subsection*{Disclosures}
The authors have no relevant financial interests in the manuscript and no other potential conflicts of interest to disclose.

\subsection*{Code and Data Availability}
The data that support the findings of this article are not publicly available due to the data policy of the XRISM Project.  Data can be requested from the author at \linkable{mizumoto-m@fukuoka-edu.ac.jp}.

\subsection* {Acknowledgments}
This work is made possible only with the significant contributions of all the XRISM
\textit{Resolve} team members, and the SHI and NEC engineers, which we greatly
appreciate. M.M. thanks Ryota Tomaru at Durham University for help of the spectral
analysis of GX 13+1. 
Part of this work was performed under the auspices of 
the Japan Society for Promotion of Science (JSPS) KAKENHI grant No.\ JP21K13958 (M.M.), Yamada Science Foundation (M.M.) and
the U.S. Department of Energy by Lawrence Livermore National Laboratory under Contract DE-AC52-07NA27344 (M.E.E.).
This paper is based on Mizumoto et al.\ {\it Proc.\ SPIE} {\bf 12181}, 121815Z (2022)

%\subsection* {Code, Data, and Materials Availability} 
%As relevant, declare the availability of computer software code, data, and/or materials used in the research results reported in the manuscript. Provide specific access information or restrictions for code, data, and materials (i.e., links to repository access addresses, and/or guidance on commercial or public access). Note: reporting in this section is required for the \textit{Journal of Biomedical Optics} and \textit{Neurophotonics}. 

%%%%% References %%%%%

\bibliography{main}   % bibliography data in report.bib
\bibliographystyle{spiejour}   % makes bibtex use spiejour.bst

%%%%% Biographies of authors %%%%%

\vspace{2ex}\noindent\textbf{Misaki Mizumoto} is a senior lecturer at University of Teacher Education Fukuoka. He received his BS, MS, and PhD degrees in astronomy from the University of Tokyo in 2013, 2015, and 2018, respectively. He was a postdoctoral fellow at Durham University since 2018 and an assistant professor at Kyoto University since 2020, with his current position beginning in 2023. His current research interests include X-ray microcalorimeter detector, active galactic nuclei, X-ray binary, and galaxy evolution.

%\vspace{1ex}
%\noindent Biographies and photographs of the other authors are not available.

\listoffigures
\listoftables

\end{spacing}
\end{document}